\documentclass[preprint]{aastex63}

\newcommand{\bmv}[1]{\mbox{\boldmath$#1$}}

\received{2020 March 5}
\revised{2020 November 23}
\accepted{2020 December 9}
\submitjournal{ApJ}

\shorttitle{Cylindrical Coordinates}
\shortauthors{Hanawa \& Matsumoto}

\begin{document}

\title{A Proper Discretization of Hydrodynamic Equations in the Cylindrical Coordinates
for Astrophysical Simulations}

\correspondingauthor{Tomoyuki Hanawa}
\email{hanawa@faculty.chiba-u.jp}

\author[0000-0002-7538-581X]{Tomoyuki Hanawa}
\affiliation{Center for Frontier Science, Chiba University, 1-33 Yayoi-cho, Inage-ku, Chiba 263-8522, Japan}

\author[0000-0002-1484-7056]{Yosuke Matsumoto}
\affiliation{Department of Physics, Graduate School of Science, Chiba University, 1-33 Yayoi-cho, Inage-ku, Chiba 263-8522, Japan}
\affiliation{Institute for Global Prominent Research, Chiba University, 1-33 Yayoi-cho, Inage-ku, Chiba 263-8522, Japan}





\begin{abstract}

Cylindrical coordinates are often used in computational fluid dynamics, in particular, 
when one considers gas flow accreting onto a central object.  Although the cylindrical coordinates
have several advantages in describing rotation, they have apparent singularity along the axis
at the coordinate origin ($ z $-axis).  This singularity introduces
difficulties in numerical simulations. First, it is difficult to reproduce the flow across the $ z $-axis.  Second, the time step is
extremely shortened by the CFL condition near the $ z $-axis because the numerical cell thereof is narrow in the azimuthal direction
for a given angular resolution.  Here, we propose a new discretization scheme to overcome these difficulties.   
In our new scheme, we consider changes in the direction of the unit vector within a cell
when evaluating the flux across each cell surface. Besides, we evaluate the source term in the
radial component of the momentum equation from the thermal and dynamic pressures working on the azimuthal cell surface.
The new scheme is designed to be free-stream preserving so that flow with uniform density, pressure, and velocity is
an exact solution of the discretized equation.
These improvements are essential to using a lower angular resolution in the innermost area and thus to 
elongate each time step.
Our examples demonstrate that the innermost circular region around the axis can be resolved by
only six numerical cells.  
We present an application to an accreting compact star surrounded by a disk, in addition to Sod shock tube and rotating outflow tests.   
\end{abstract}



\section{Introduction \label{sec:intro}}

The cylindrical coordinates, $ (r, \varphi, z) $, are appropriate for describing accretion onto a compact
object and outflow from the disk surrounding it.
They can describe centrifugal balance, angular momentum transfer, and vertical structure along the
rotation axis in a simple form. Thus, they can be adopted in the numerical simulations of
accreting protostars, neutron stars, and black holes.  Although the simulations thus far have given us
useful insights regarding these astronomical objects, they are often associated with
technical problems.  First, they often show unphysical disturbances around the $ z $-axis, 
hindering wave propagation and nonaxisymmetric structures of physical origin.
Second, the time step, $ \Delta t $, is restricted to be much shorter than dynamic time scales in the outer regions because the CFL condition \cite[see, e.g.,][]{hirsch90} is determined by the azimuthal innermost cell width, $ r _{\rm min} \Delta \varphi $.  The short time step wastes our computational resources and is an obstacle
for pursuing a long time evolution.  If we can remove these disadvantages, the cylindrical coordinates
could be more useful for numerical simulations.  

Thus far, many researchers have proposed numerical methods to alleviate the disadvantages 
mentioned.   
The simplest method is to set an inner boundary at a certain radius and
to remove the region inside the boundary close to the computation domain \cite[see, e.g.,][]{hawley13}.
Similarly, conical regions around the $ z $-axis are often omitted in the 
numerical simulations performed in the spherical polar coordinates \cite[see, e.g.,][]{suzuki15,mizuta18}.
However, we cannot apply this method if we are interested in outflows along the $ z $-axis.
Another simple cure is averaging or smoothing. A small-scale disturbance is
removed if we replace the density, velocity, and pressure in a numerical cell close to
the origin ($ r = 0 $) with the averages of the nearby cells.  
At the same time, the averaging
allows us to take a longer time step.   
A simple averaging results in numerical damping.  We need a more sophisticated treatments such as cell merging \citep{asaithambi17} or ring averaging \citep{zhang19} to avoid additional numerical viscosity.\footnote{Both the methods are applied not to the velocity components ($ v _r $ and $ v _\varphi$) in the cylindrical coordinates, but to th in the Cartesian coordinates.}  Although a numerical scheme of a higher-order spatial accuracy 
may remove the difficulty, such a scheme has not been provided yet.  

In this paper, we propose a new numerical scheme to solve the hydrodynamic equations
in the cylindrical coordinates to overcome such difficulties inherent in the cylindrical coordinates.   
Our proposed method is based on the monotonic upwind scheme for conservation law (MUSCL).   
First, we evaluate the centrifugal force not at the cell center but on the cell surface.
This treatment can be understood from the idea that the centrifugal force is a part of the dynamic
pressure working on the cell surface in the $ \varphi $-direction.
Second, we consider changes in the
unit vectors in the $ r $- and $ \varphi $-directions along the $ \varphi $-coordinates when
we reconstruct the right- and left-hand-side states of a numerical cell boundary.  
We evaluate the velocity components normal and tangential to the cell boundary from
those in the nearby cells.  Note that the component normal to the cell surface is not the
$ \varphi $-component at the cell center because the $ \varphi $-directions at the cell
center are different than those on the cell surface.  Third, we introduce a discretized form of
hydrodynamic equations so a uniform flow is an exact solution of it.  
When the density, pressure, and velocity in the Cartesian coordinates are uniform,
the flow should be stationary.  However, such a uniform flow is not an exact
stationary solution if we adopt a naive discretization form of the equations in the cylindrical coordinates.
These three improvements suppress the appearance of irregularities
in the numerical solutions obtained by the current standard scheme.
At the same time, they allow us to take a large azimuthal cell width, $ \Delta \varphi = \pi / 3 $,
in the inner region adjacent to the $ z $-axis.  Fourth, we employ a variable angular resolution
to achieve a better aspect ratio, $ \Delta r \approx r \Delta \varphi $, where $ \Delta r $
denotes the radial cell width.  The coarse--fine interface is treated as in the adaptive mesh refinement (AMR).
The last improvement resolves the short time step problems,
while the other three restore numerical stability.  Lowering the angular resolution around the
axis itself has already been adopted in some numerical schemes \cite[see, e.g.,][]{liska18}, and we compare our scheme
with the previous ones.   Although some early results are shown in \cite{hanawa20},
this paper contains some new improvements. 

This paper is organized as follows.   In the first half of \S 2, we summarize the current 
standard method to solve the hydrodynamic equation based on MUSCL.
In the second half of \S2, we inspect the hydrodynamic equations to  
elucidate the source of numerical troubles.   In \S3, we show our numerical procedure
to solve the hydrodynamic equations.  In \S 4, we show some numerical tests
to validate our scheme.   They include uniform flows, the Sod shock tube problem,
rotating and expanding gas sphere, and gas-disk interaction around a compact star.  In \S 5, we review our method and discuss
possible extensions of our scheme to magnetohydrodynamics and hydrodynamic simulations in 
other curvilinear coordinates such as the spherical coordinates.
\section{Critical Inspection of the Current Standard Method}

\subsection{Current Standard Approach}

In this section, we review the monotonic upwind scheme for conservation law, MUSCL,
because our new scheme is based on it. 
When applying a MUSCL approach, we express the hydrodynamic equations in the
conservation form,
\begin{eqnarray}
\frac{\partial \bmv{U}}{\partial t} + \frac{\partial \bmv{F}_r}{\partial r} 
+ \frac{\partial \bmv{F} _\varphi}{\partial \varphi} + \frac{\partial \bmv{F} _z}{\partial z} 
& = & \bmv{S},  \label{hydro1}
\end{eqnarray}
where
\begin{eqnarray}
\bmv{U} = 
\left[ \begin{array}{c} r \rho \\ r \rho v _r \\ r ^2 \rho v _\varphi \\ r \rho v _z \\ r \rho E \end{array} \right],
\hskip 5pt
\bmv{F} _r = 
\left[\begin{array}{c} r \rho v _r \\ r \left( \rho v _r {}^2 + P\right) \\ r ^2 \rho v _r v _\varphi \\ r \rho v _r v _z \\
r \rho v _r H \end{array} \right], \hskip 5pt
\bmv{F} _\varphi = 
\left[ \begin{array}{c} \rho v _\varphi \\ \rho v _\varphi v _r \\  r \left( \rho v _\varphi {}^2 + P \right) \\ 
\rho v _\varphi v _z \\
\rho v _\varphi H \end{array} \right], \nonumber \\
\bmv{F} _z = 
\left[ \begin{array}{c} r \rho v _z \\ r \rho v _z v _r \\  r ^2 \rho v _z v _\varphi \\ 
r \left( \rho v _z {}^2 + P \right) \\ r \rho v _z H \end{array} \right], \hskip 5pt
\bmv{S} = 
\left[ \begin{array}{c} 0 \\ r \rho g _r + {\rho v _\varphi {}^2 + P} \\
r ^2 \rho g _\varphi \\ r \rho g _z \\ r \rho \bmv{g} \cdot \bmv{v} \end{array} \right].  
\label{hydro2}
\end{eqnarray}
Here, $ \rho $ and $ P $ denote the gas density and thermal pressure, respectively.
The symbols, $ v _r $, $ v _\varphi $, and $ v _z $ denote the $ r $-, $ \varphi $-, and
$ z $-components of the gas velocity, respectively, while $ g _r $, $ g _\varphi $, and $ g _z $
denote the $ r $-, $ \varphi $-, and $ z $-components of the gravity, respectively.
The symbols $ E $ and $ H $ denote the specific energy and specific enthalpy, 
respectively, and are expressed as
\begin{eqnarray}
E & = & \frac{v _r {}^2 + v _\varphi {} ^2 + v _z {}^2}{2} + \frac{P}{\left( \gamma -1 \right) \rho} , \\
H & = & E + \frac{P}{\rho},
\end{eqnarray}
for an ideal gas having the specific heat ratio, $ \gamma $.  

The first and last components of Equation (\ref{hydro1}) denote the mass and energy 
conservation, respectively, while the other components denote the equation of motion.
The third and fourth components correspond to the conservation
of angular momentum and that of linear momentum in the $ z $-direction, respectively.
The second component denotes the radial component of the equation of motion but not 
the conservation of radial momentum. This decomposition of the equation of motion is appropriate for the
simulations of gas accretion onto a compact object because the angular momentum is a key 
parameter.   Thus, Equation (\ref{hydro1}) is a standard form used to denote the hydrodynamic
equations in the cylindrical coordinates in astrophysics \cite[see, e.g.,][]{mignone07,skinner10}.  However,
another set of equations in which each component denotes the conservation of
linear momentum in the Cartesian coordinates is often used in other fields of computational 
fluid dynamics including aeronautics \cite[see, e.g.,][]{vinokur89}. 

In the current standard MUSCL approach, we integrate Equation (\ref{hydro1}) over
the small volume of $ r _i - \Delta r _i/2  < r < r _i + \Delta r _i /2 $,
$ \varphi _j - \Delta \varphi _j/2 < \varphi < \varphi _j + \Delta \varphi _j /2 $, and
$ z  _k - \Delta z _k /2 < z < z _k + \Delta z _k / 2 $, where $ (r _i, \varphi _j, z _k ) $ 
and $ (\Delta r _i, \Delta \varphi _j, \Delta z _k ) $ denote the central position 
and width of the small volume numbered by integers of 
$ i $, $ j $, and $ k $, respectively.  
We call this small volume a numerical cell.   The numerical cell $(i,j,k) $ has
the following volume,
\begin{eqnarray}
\Delta V _{i,j,k} & = & r _i \Delta r _i \Delta \varphi _j \Delta z _k ,
\end{eqnarray}
and is bounded by 
the six cell surfaces,
\begin{eqnarray}
\Delta S _{r,i\pm1/2,j,k} & = & \left( r _i \pm \frac{\Delta r _i}{2} \right)  \Delta \varphi _j \Delta z _k , \\
\Delta S _{\varphi,i,j\pm1/2,k} & = & \Delta r _i \Delta z _k , \\
\Delta S _{z,i,j,k\pm1/2} & = & r _i  \Delta r _i \Delta \varphi _j .
\end{eqnarray}
We evaluate $ \bmv{U} $ and $ \bmv{S} $ as the cell volume average, while 
we evaluate $ \bmv{F} _r $, $ \bmv{F} _\varphi $, and $ \bmv{F} _z $ as the cell surface average.
In the following we use the term \lq \lq standard\rq \rq to mean that the source term 
$ (P + \rho v _\varphi {} ^2)$ is evaluated as the cell volume average.
For later convenience, we define the modified volume and surface elements,
\begin{eqnarray}
\Delta V _{i,j,k} ^\prime & = & r _i {} ^{-1} \Delta V _{i,j,k} \, = \, \Delta r _i \Delta \varphi _j \Delta z _k , \\
\Delta S _{r,i+1/2,j,k} ^\prime & =& r _{i+1/2} {} ^{-1} \Delta  S _{r,i+1/2,j,k} \, = \, \Delta \varphi _j \Delta z _k ,\\
\Delta S _{z,i,j,k+1/2} ^\prime & = & r _i {} ^{-1} \Delta S _{z,i,j,k+1/2} \, = \Delta r _i \Delta \varphi _j .
\end{eqnarray}
Then, we obtain the spatially discretized form of the hydrodynamic equations,
\begin{eqnarray}
\frac{\partial \bmv{U} _{i,j,k}}{\partial t}  \Delta V _{i,j,k} ^\prime   +
\bmv{F} _{r,i+1/2,j,k} \Delta S _{r,i+1/2,j,k} ^\prime - \bmv{F} _{r,i-1/2,j,k} \Delta S _{r,i-1/2,j,k} ^\prime
+ \bmv{F} _{\varphi,i,j+1/2,k} \Delta S _{\varphi,i,j+1/2,k} \nonumber \\
 - \bmv{F} _{\varphi,i,j-1/2,k} \Delta S _{\varphi,i,j-1/2,k}
 + \bmv{F} _{z,i,j,k+1/2} \Delta S _{z,i,j,k+1/2} ^\prime
  - \bmv{F} _{z,i,j,k-1/2} \Delta S _{z,i,j,k-1/2} ^\prime = \bmv{S} _{i,j,k} \Delta V _{i,j,k} ^\prime  , 
  \label{Dhydro1}
\end{eqnarray} 
where the suffixes of  $ \bmv{U} $, $ \bmv{F} _r $, $ \bmv{F} _\varphi $,  $ \bmv{F} _z $, and
$ \bmv{S} $ specify the cell volume or cell surface where the values are evaluated.  
The modified volume and surface elements compensate the multiplication factor, $ r $, in
$ \bmv{U} $, $ \bmv{F} _r $,  $ \bmv{F} _\varphi $, $ \bmv{F} _z$, and $ \bmv{S} $.

The accuracy of a solution depends on the method of evaluating the flux across a cell
and that of time integration.  The accuracy in space depends on the former, while that
in time depends on the latter.

In the MUSCL approach, we evaluate the flux across a cell surface from the density, 
velocity, and pressure in the cells aligned normally to the cell surface.  First, we evaluate 
the density, velocity, and pressure on the left- and right-hand sides of the cell surface 
by reconstructing (interpolating or extrapolating) these variables.  
The accuracy in space depends on the choice of reconstruction, and  
a higher-order reconstruction results in higher spatial accuracy.   However, reconstruction
may produce artificial extrema when the slope of a variable changes around the
cell surface. Because such artificial extrema have origins of numerical oscillations, we 
set a limit on the reconstruction to keep monotonicity. The accuracy and monotonicity are
often trade-offs, and there are various methods for data reconstruction.  Second, we 
evaluate the flux across the cell surface from the left- and right-hand-side states
using an approximate Riemann solution.  As summarized in \cite{toro09}, there are
several Riemann solvers for hydrodynamic equations.   In general, less diffusive
solvers are more accurate but less stable.  

The accuracy in time is limited by the number of steps used to update the density,
velocity, and pressure.  If we employ a two-step
method, the accuracy is second-order in time.   If we take the third-order Runge--Kutta 
method, the temporal accuracy is third order.

\subsection{Critical Inspection of the Current Standard Methods}

In this subsection, we inspect the scheme described in the previous subsection to improve it.

First, we focus our attention on the source term, $ \bmv{S} $.  In the absence of gravity,
the source term, $ P + \rho v _\varphi {}^2 $, remains in the second component in $ \bmv{S} $ 
in the cylindrical coordinates [see Equation (\ref{hydro2})].   To the best of our knowledge,
this term is evaluated as the cell average in the simulations thus far \citep[see, e.g.,][]{skinner10,mignone14}.
However, we surmise that we should evaluate it on the cell surface because it coincides with the total pressure working
on the cell surface in the $\varphi $-direction.  

\begin{figure}[ht]
\epsscale{0.5}
\plotone{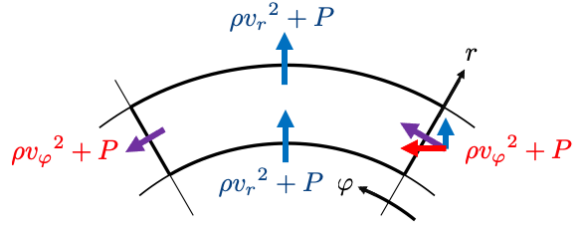}
\epsscale{1.0}
\caption{Illustration of the total pressure working on the cell surfaces in
the cylindrical coordinates.  A fraction of the total pressure working on the azimuthal cell surface
has a radial component.  \label{ram_pressure}} 
\end{figure}

Figure \ref{ram_pressure} illustrates the idea intuitively.   A numerical cell is arc-like on the $r\varphi $-plane.
The left and right cell surfaces are inclined by $ \Delta \varphi / 2 $ with respect to the cell
center, $ \varphi = \varphi _j $, and the total pressure works in the $ \varphi $-direction on the cell surface.   
However, the dynamical pressure contains not only the $\varphi$-component of the momentum ($v _\varphi$) flux into the $\varphi$-direction but also that of the $r $-component of the
momentum ($v _r $) flux in the $ \varphi $-direction. Note that the unit vector in the $\varphi$-direction is
$ \bmv{e} _\varphi (\varphi _j) $ at the cell center but
$ \bmv{e} _\varphi (\varphi _j \pm \Delta \varphi/2 )
= \cos \Delta \varphi/2\ \bmv{e}_\varphi \mp \sin \Delta \varphi/2\ \bmv{e} _r$
on the azimuthal cell surface.  Thus, the $ \varphi $-component of the
momentum at the cell surface contains the $ r $-component of momentum
when evaluated at the cell center.
This seemingly
curious interpretation comes from the curvature of the coordinates.  While the $ r $-direction
changes with the change in $ \varphi $, the radial momentum flux from
each cell surface is evaluated to be
\begin{eqnarray}
\left(  P  + \rho v _\varphi {}^2 \right) \sin \frac{\Delta \varphi}{2} \Delta r \Delta z 
\simeq \left(  P + \rho \varphi _{} ^2 \right) \frac{\Delta \varphi}{2} \Delta r \Delta z .
\end{eqnarray}
The right-hand side coincides with the source term except for the factor, $ 1/2 $.
The other half is provided from the other cell surface.
This coincidence has been known for
decades in aeronautical computational fluid dynamics \citep[see, e.g.][]{vinokur74}, and we will discuss another argument supporting this
idea in \S 5.  

We can reword above argument for evaluating the source term on 
the azimuthal cell surfaces in terms of mathematics.  
Note that the equation of momentum conservation
has a vector form, and the equation is expressed as follows,
\begin{eqnarray}
\frac{\partial}{\partial t} \left( \rho \bmv{v} \right) + \bmv{\nabla} \cdot \bmv{T}  =  \rho \bmv{g} , \\
\bmv{T} = \rho \bmv{v}\bmv{v} + P \bmv{I}, 
\label{hydroV}
\end{eqnarray}
where $ \bmv{T} $ and $ \bmv{I} $ denote momentum flux and unit tensors, respectively. 
By integrating Equation (\ref{hydroV}) over the volume, we obtain the following,
\begin{eqnarray}
\frac{\partial}{\partial t} \int\int\int  _{V} \rho \bmv{v} dV + \int\int _{\partial V} \bmv{T} \cdot d\bmv{S}
& = & \int\int\int _{V} \rho \bmv{g} dV , \label{hydroV2}
\end{eqnarray}
where the second term in the left-hand side denotes the surface integral of the momentum
flux tensor. The finite volume method should be based on this integral form.
Equation (\ref{hydroV2}) indicates that we need not include any extra
source term if we evaluate the surface integral properly.

Second, we observe the method to evaluate the velocity in the left- and right-hand sides of
a cell surface.  When computing the numerical flux across a cell boundary, we need the
normal and tangential components of the velocity, $ v _{\rm n} $, $ v _{\rm t1} $, and $ v _{\rm t2} $.  
As mentioned,
the azimuthal cell boundary is slightly inclined with respect to the radial direction at the cell
center.  By considering the inclination, we should evaluate them by
\begin{eqnarray}
v _{\rm n} \left(\varphi; \varphi _{j+1/2} \right) & = & v _\varphi \left(\varphi \right)  
\cos \left( \varphi - \varphi _{j+1/2} \right)
+ v _r \left( \varphi \right) \sin \left( \varphi - \varphi _{j+1/2} \right) ,  \label{vn} \\
v _{\rm t1} \left(\varphi; \varphi _{j+1/2} \right) & = & v _r \left(\varphi \right)  
\cos \left( \varphi - \varphi _{j+1/2} \right)
- v _\varphi \left( \varphi \right) \sin \left( \varphi - \varphi _{j+1/2} \right) , \label{vt1} \\
v _{\rm t2} \left(\varphi; \varphi _{j+1/2} \right) & = & v _z \left( \varphi \right),
\end{eqnarray}
where $ \varphi $ and $ \varphi _{j+1/2} $ denote the azimuthal angle at the cell center and that on
the cell surface, respectively. 
We should reconstruct these normal and tangential components rather than $ v _r $ and $ v _\varphi $ 
to evaluate the left-
and right-hand sides.   Otherwise, we cannot set an appropriate limit on the data reconstruction.
Suppose the velocity is uniform in the Cartesian coordinates. Then, the normal and tangential
components are constant and do not depend on $ \varphi $, and accordingly, the profiles 
should be reconstructed without a slope limiter.  However, both $ v _r $ and $ v _\varphi $ depend on $ \varphi $
and may have an extremum in the $ \varphi $-direction. Then, the current standard
method sets a limit on the slope and produces an artificial extremum in $ v _{\rm n} $
and $ v _{\rm t1} $.  It is quite ironic that the limiter intended to remove an artificial 
extremum produces another artificial extremum. 

The presented procedure is equivalent to using the covariant derivative of the velocity in the
data reconstruction \citep[see, e.g.][]{mitra09}.  The covariant derivative
should include higher-order terms either when the data reconstruction is of a higher-order
accuracy or when the angular resolution is low.  It is easier to use
Equations (\ref{vn}) and (\ref{vt1}) than to compute derivatives of the Christoffel
symbols. 

Third, we inspect free-stream preservation, i.e., whether a uniform flow is a stationary solution 
of the system equation to be solved.   When the density, velocity, and pressure are uniform in the domain,
the flow is uniform and stationary in the absence of gravity.  This condition should be satisfied in the discretized space and time in arbitrary coordinate systems. In contrast, a proper discretization for the stationary condition can be straightforwardly obtained in the Cartesian coordinates, and it is not trivial in the cylindrical coordinates.  This problem implies
that the current standard discretization in space produces a spurious inertial force.  
In short, we conclude that the current standard discretization does not preserve the 
free stream. The differential operator, $ \bmv{\nabla} $, should be discretized properly
to ensure free-stream preservation \citep[see, e.g.,][for more details on the free-stream preservation]{vinokur89,wongwathanarat16}.

Fourth, we want to elongate the time step to follow a long time evolution at
a relatively small computational cost.  When the angular resolution, $ \Delta \varphi $,
is uniform, the time step is most tightly constrained in the innermost cells by the CFL
condition because the azimuthal width is much smaller than the radial width, $ r  \Delta \varphi 
\ll \Delta r $ in the innermost cells.  Thus, we can take a longer time step if the angular resolution is
reduced near the $ z $-axis ($ r = 0 $).    The CFL condition is less tight when the cell
has nearly the same width in the $ r $-, $ \varphi $-, and $ z $-directions.
Thus, the aspect ratio, $ r _i \Delta \varphi _j / \Delta r $
and $ \Delta z _k / \Delta r _i $, should be close to unity in the optimal numerical grid.
It should have  $ \Delta \varphi = \pi / 3 $ in the innermost
cells around the $ z $-axis because they are similar to the equilateral triangle.  
When the aspect ratio is close to unity, the numerical diffusivity is nearly isotropic.
This is another merit of adjusting the aspect ratio close to unity.

Although these four issues are essentially independent, the last one is related to the other three.  
We can lower the angular resolution without
sacrificing accuracy and stability only when we can resolve the other issues at a low angular 
resolution.  

\section{Numerical Methods \label{sec:method}}


In this section, we introduce our recipes to solve the four issues raised in \S\S 2.2.

\begin{figure}
\plottwo{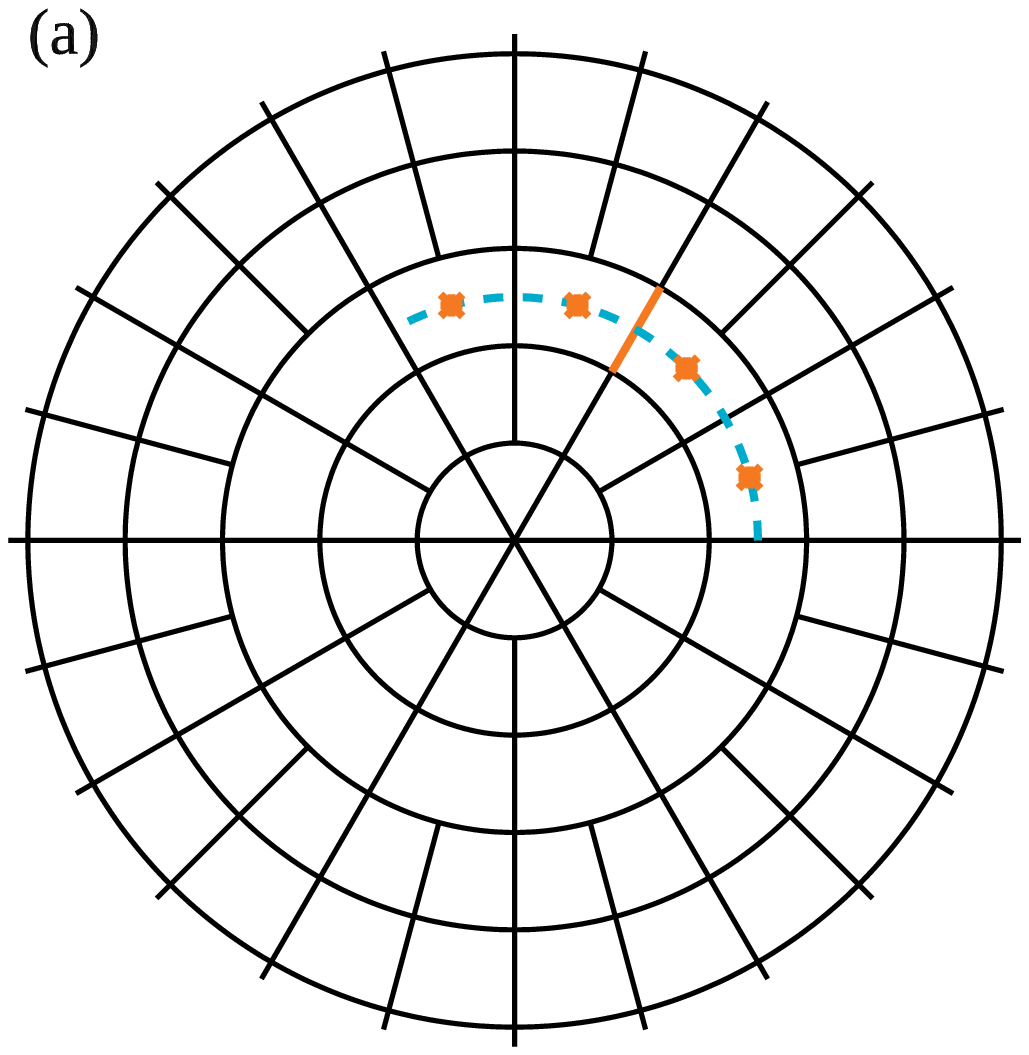}{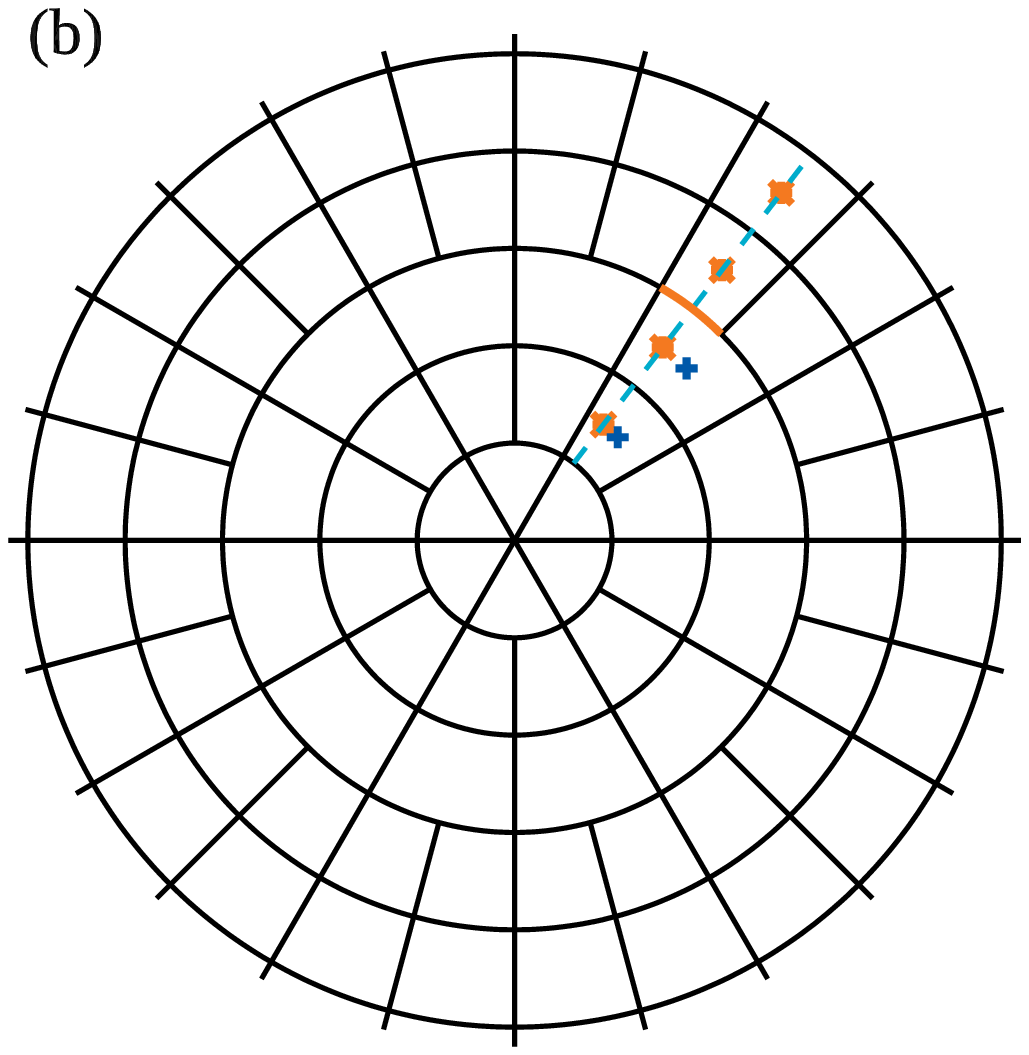}
\caption{These two panels show the numerical grids used in our simulations.
The solid black curves denote the cell boundaries in the $ r $-direction while
the solid black lines denote those in the $ \varphi $-direction. The left and right panels
illustrate the data reconstruction on the $ \varphi $- and $ r $-directions, respectively,
for computing the numerical flux across the red cell boundaries.   We reconstruct the
data on the left and right sides of the boundary along the broken blue curves and lines.\label{AMR}}
\end{figure}

First, we evaluate the source term as
\begin{eqnarray}
\bmv{S} _{i,j,k} & = & \frac{1}{2} \left[ 
\begin{array}{c}
0 \\ \rho _{i,j,k} \left( r _{i+1/2} g _{r,i+1/2,j,k} + r _{i-1/2} g _{r,i-1/2,j,k} \right) 
+ \left(  P + \rho v _\varphi {}^2 \right) _{i,j+1/2,k} + \left(  P + \rho v _\varphi {}^2 \right) _{i,j-1/2,k} \\
 \rho _{i,j,k}  r _i {}^2 \left( g _{\varphi,i,j+1/2,k} + g _{\varphi,i,j-1/2,k} \right) \\
 \rho _{i,j,k}  r _i \left( g _{z,i,j,k+1/2} + g _{z,i,j,k-1/2} \right) \\
\left( r \rho v _r \right) _{i+1/2,j,k} g _{r,i+1/2,j,k}  + \left( r \rho v _r \right) _{i-1/2,j,k} g _{r,i-1/2,j,k}  
+ \left( r \rho v _\varphi \right) _{i,j+1/2,k} g _{\varphi,i,j+1/2,k} \hskip25pt \; \\
\, \hskip 15pt + \left( r \rho v _\varphi \right) _{i,j-1/2,k} g _{\varphi,i,j-1/2,k}
+ \left( r \rho v _z \right) _{i,j,k+1/2} g _{z,i,j,k+1/2}
+ \left( r \rho v _z \right) _{i,j,k-1/2} g _{z,i,j,k-1/2}
\end{array} \right], \label{modified_source}
\end{eqnarray}
where $ g _r $, $ g _\varphi $, and $ g _z $ denote the $ r $-, $\varphi $-, and $ z $-components
of the gravity.   They are evaluated on the cell surface designated by the suffixes and related
to the gravitational potential, $ \Phi $, by
\begin{eqnarray}
g _{r,i+1/2,j,k} & = & - \frac{2 \left( \Phi _{i+1,j,k} - \Phi _{i,j,k}\right)}{\Delta r _{i+1} + \Delta r _i}, \\
g _{\varphi,i+1/2,j,k} & = & - \frac{2 \left( \Phi _{i,j+1,k} - \Phi _{i,j,k}\right)}{
r _i \left(\Delta \varphi _{j+1} + \Delta \varphi _j \right)}, \\
g _{z,i,j,k+1/2} & = & - \frac{2 \left( \Phi _{i,j,k+1} - \Phi _{i,j,k} \right)}{\Delta z_{k+1} + \Delta z _k}.
\end{eqnarray}
If we use this form of the source term, then, the total energy of the system is conserved (see, e.g., 
\cite{hanawa19}).  This modified source term realizes the first issue raised in \S\S 2.2.  
Equation (\ref{modified_source}) looks similar to the trapezoidal rule of \citet{mignone14}, which suggests to approximate the source term by the average of the cell surface values in the radial direction.  However, Equation (\ref{modified_source}) takes the average of the dynamical pressure on the azimuthal cell surfaces. These two averages are essentially different.
Second, we employ the numerical grid shown in Figure \ref{AMR} to realize the fourth issue
raised in \S \S 2.2.   The black curves and lines denote the cell boundaries on the
$ r-\varphi $ plane.  The central circle ($i = 1$) is divided into six numerical cells while the adjacent ring
($ i = 2 $)  is divided into 12 numerical cells.   The angular resolution is set to be a function of $ r $ so that
\begin{eqnarray}
\frac{\Delta r _i}{\sqrt{2}  r _i} < \Delta \varphi _i = \frac{\pi}{3 \times 2 ^n} < 
\frac{\sqrt{2} \Delta r _i}{r _i} ,   \label{grid}
\end{eqnarray}
where $ n $ denotes a positive integer less than $ n _{\rm max} $ and a function of $ i $.  
The azimuthal cell boundaries are aligned on $ \varphi = 0 $ throughout the computation region to synchronize 
them.  The radial cell width is
set to be constant up to the $ 2 ^{n_{\rm max}} $-th cell, and beyond which, it increases
by a factor $ 1 + 2 ^{-n_{\rm max}} $ each with increased cell number, $ i $,
in the $ r $-direction to keep the aspect ratio, $  r _i \Delta \varphi _i /  \Delta r _i$,  close to unity.  
The cell width in the $ z $-direction, $ \Delta z _k $, is set to be constant in
the region around the mid plane, $ z = 0 $, while it increases as $ |z | $ increases in the region far
from $ z = 0 $.

This numerical grid is similar to that of the block-structured AMR method adopted
in the hydrodynamic codes, Athena++ \citep{stone19}, Pluto \citep{mignone07}, Ramses \citep{teyssier02}, SFUMATO \citep{matsumoto07}, and so on.
At some radii, numerical cells contact with two inner ones in the $ r $-directions.  We treat these
coarse--fine (wide--narrow) interfaces according to AMR. The numerical flux into the coarse cell 
is equated with the sum of the numerical flux into the fine,
\begin{eqnarray}
\bmv{F} _{r,i+1/2,j,k} ^{\rm (c)} \Delta S _{r,i+1/2,j,k} ^{\rm (c)} 
& = & \bmv{F} _{r,i+1/2,2j-1,k} ^{\rm (f)} \Delta S _{r,i+1/2,2j-1,k} ^{\rm (f)} + 
\bmv{F} _{r,i+1/2,2j,k} ^{\rm (f)} \Delta S _{r,i+1/2,2j} ^{\rm (f)} , \\
\Delta S _{r,i+1/2,j,k} ^{\rm (c)} & = & \Delta S _{r,i+1/2,2j-1,k} ^{\rm (f)} +  
\Delta S _{r,i+1/2,2j,k} ^{\rm (f) }, 
\label{AMRr}
\end{eqnarray}
in order to keep the conservation form.   Here, the superscripts (c) and (f) represent
coarse and fine, respectively.  Note that the cell number in the $ \varphi $-direction should
be arranged to match the coarse and fine cell surfaces.

Because we use the minmod limiter to preserve the monotonicity, we use two contiguous zones
in both sides of the cell boundary where we evaluate the numerical flux.   If one of them is
subdivided into more than two cells, we subdivide the cell boundary to fit with the smallest
tangential cell width.   For example, we divide the radial boundary at $ r  = r _{5/2} $ into
two, although the azimuthal cell width is $ \Delta \varphi = \pi / 6 $ (see Figure \ref{AMR}).
The cell boundary width should be equal to that in the cell at $ r = 4 $, since the density, 
velocity, and pressure thereof are used in the data reconstruction.   This subdivision is 
a minor change from the  AMR.

Third, we consider the change in the unit vectors, $ \bmv{e} _r $ and $ \bmv{e} _\varphi $,
along the $\varphi$-coordinates (the blue dashed curve in the left panel of Figure \ref{AMR})
when evaluating the left- and right-hand sides of a numerical
cell surface (marked orange in the left panel of Figure \ref{AMR}).    
When evaluating the left- and right-hand sides of an azimuthal cell surface, we
transform $ v _r $ and $ v _\varphi $ into $ v _{\rm n} $ and $ v _{\rm t1} $ according to
Equations (\ref{vn}) and (\ref{vt1}) and then apply the minmod limiter to $ v _{\rm n} $,
$ v _{\rm t1} $, and $ v _z $.   The density and pressure thereof are evaluated by the
standard procedure.

We also consider the change in the unit vectors at the coarse--fine interface in
the radial direction, and the right panel of Figure \ref{AMR} illustrates this situation.
The orange arc denotes the cell surface of interest.  The blue
dashed line is normal to the orange arc and goes through the midpoint.
It goes through the centers of the outer cells but not those of the inner ones. 
We evaluate the density and pressure on the blue line to be equal to those
at the cell center.   We also evaluate the velocity on the blue line to be equal
to that at the cell center but while considering the changes in $ \bmv{e} _r $
and $ \bmv{e} _\varphi $.  In other words, we do not consider the gradient of 
density, velocity, and pressure inside the coarse cell, but we do consider the changes in the unit vectors.
Even when the velocity vector is uniform, it has different $ r $- and $ \varphi $-
components between the cell center and the cell boundary.

Fourth, we modify the surface elements to achieve the free-stream preservation
so that a uniform flow is an exact solution of the discretized hydrodynamic equations.
When the velocity is uniform, the $ r $- and $ \varphi $-components depend on $ \varphi $,
\begin{eqnarray}
v _r (r, \varphi, z) & = & v _{x0} \cos \varphi + v _{y0} \sin \varphi ,  \label{vr1} \\
v _\varphi (r, \varphi, z ) & = & - v _{x0} \sin \varphi + v _{y0} \cos \varphi , \label{vp1}
\end{eqnarray}
where $ v _{x0} $ and $ v _{y0} $ denote the $ x $- and $ y $-components of the velocity
in the uniform flow, respectively.   We seek the condition that Equations (\ref{vr1}) and
(\ref{vp1}) are the exact stationary solution of Equations (\ref{Dhydro1}) by modifying the surface 
elements.  We can modify either $ \Delta S _{r,i+1/2,j,k} $ or $ \Delta S _{\varphi,i,j+1/2,k} $
to make the uniform flow the exact solution.  However, it is easier to modify 
$ \Delta S _{\varphi,i,j+1/2,k} $ than $ \Delta S _{r,i+1/2,j,k} ^\prime $ because the latter should
also meet the condition (\ref{AMRr}).   We modify the surface elements to multiply the 
correction factor, $ f $, listed in Table \ref{Scorrection}.    The correction factor is common for
the mass conservation, the $ z $-component of the equation of motion, and the energy 
conservation, and is denoted by $ f _{\rho} $.   We multiply the correction factors,
$ f _{r} $ and $ f _{\varphi} $, for the $ r $- and $ \varphi $-components of the equation of motion,
respectively.
The correction factor depends also on whether the radial surface elements are subdivided or not.
The numerical grid used in this paper has three types of cells:
\begin{description}
\item[(A)] Neither the inner nor outer radial surface element is divided into two: \\
$ \Delta \varphi (r _{i-1/2}) = \Delta \varphi (r _i) = \Delta \varphi (r _{i+1/2}) $.
\item[(B)] The outer radial surface element is divided into two while the inner one is not: \\
$ \Delta \varphi (r _{i-1/2}) = \Delta \varphi (r _i) = 2 \Delta \varphi (r _{i+1/2}) $.
\item[(C)] Both the inner and outer radial surface elements are divided into two: \\
$ 2 \Delta \varphi (r _{i-1/2}) = \Delta \varphi (r _i) = 2 \Delta \varphi (r _{i+1/2}) $.
\end{description}
All these correction factors converge to unity  
in the limit of $ \Delta \varphi \rightarrow 0 $.  The deviation of a correction factor from unity is
proportional to $ \left( \Delta \varphi \right) ^2 $ in the limit.  
Thus, our discretization is of the second-order accuracy in space.  Note that these correction factors
are close to unity even in the case of a low angular resolution.
They are $ f _\rho = 1.0115 $, $ f _r = 0.8259 $, and $ f _\varphi = 1.0472 $ for the innermost
cells around the $ z $-axis  ($ i = 1 $) and
$ f _\rho = 0.9942 $, $ f _r = 0.9414 $, and $ f _\varphi = 0.9996 $ for the second
innermost ones ($ i = 2 $).

\begin{table}
\begin{center}
\begin{tabular}{lccc}
\hline
cell type & (A) & (B) & (C) \\
\hline
$ f _\rho $ & $ \displaystyle \frac{\Delta \varphi}{2} \left(  \sin \frac{\Delta \varphi}{2} \right) ^{-1} $ & 
$ \displaystyle \frac{\Delta \varphi}{2} \left(  \sin \frac{\Delta \varphi}{2} \right) ^{-1}
\left[ 1 - \frac{2 r _{i+1/2}}{\Delta r _i} \sin ^2 \frac{\Delta \varphi}{8} \right] $ & 
$ \displaystyle \frac{\Delta \varphi}{4} \left(  \tan \frac{\Delta \varphi}{4} \right) ^{-1}  $ \\
$ f _r $ & $ \displaystyle \left( \frac{\Delta \varphi}{2} \right) 
\left(  \tan \frac{\Delta \varphi}{2} \right) ^{-1} $ &
$ \displaystyle \frac{\Delta \varphi}{\sin \Delta \varphi}
\left( 1 - \frac{r _{i+1/2} + \Delta r _i}{\Delta r _i}
\sin ^2 \frac{\Delta \varphi}{4} \right) $ & 
$ \displaystyle 
\frac{\Delta \varphi}{\sin \Delta \varphi}
\left[ 1 - 2 \sin ^2 \frac{\Delta \varphi}{4} \right]$ \\
$ f _\varphi  $ & $ \displaystyle \frac{\Delta \varphi}
{\sin \Delta \varphi} $ &
  $ \displaystyle \frac{\Delta \varphi}{\sin \Delta \varphi} \left[ 1 
 - \frac{r _{i+1/2} ^2}{r _i \Delta r _i} \sin ^2 \frac{\Delta \varphi}{4} \right] $  &
 $ \displaystyle \frac{\Delta \varphi}{2} \left( \sin \frac{\Delta \varphi}{2} \right) ^{-1} $  \\
\hline
\end{tabular}
\end{center}
\caption{
The correction factor to the azimuthal surface elements.   The symbol
$ \Delta \varphi $ denotes the angular cell width, while $ r _{i+1/2} $ and 
$ \Delta r _i $ denote the outer radius of the cell and the radial cell width,
respectively.  See the text for the cell types (A)--(C). \label{Scorrection}}
\end{table}
 
\section{Numerical Examples}

We assume that the gas is an ideal one having the specific heat ratio, $ \gamma $.   It is set to
 be $ \gamma = 5/3 $ in all the test problems. 
 
 We use the two-step Runge--Kutta method to integrate in time.   
 The time step is specified by the following,
 \begin{eqnarray}
\Delta t & = & {\rm CFL} \frac{\min \left[ \Delta r _{\rm min} ,  (r \Delta \varphi ) _{\rm min} \right]}
{\max (| \bmv{v} | + c _s |)}, \label{CFL}
\end{eqnarray}
where CFL and $ c _s $ denote the CFL number and the sound speed.    Equation
(\ref{CFL}) does not consider the vertical width
because its minimum is set equal to the minimum radial width, $ \Delta z _{\rm min} =
\Delta r _{\rm min} = 0.1 $ unless otherwise noted.  
The CFL number is set to be either $ {\rm CFL} = 0.3 $ or 0.5.  Note that
the minimum azimuthal width, $ (r \Delta \varphi ) _{\rm min} $, is smaller
than the radial width, $ \Delta r _{\rm min} $, in our grid. Thus, the time
step is constrained by the azimuthal width.  

We use the fixed boundary in the $ r $-direction, while we apply either the fixed or periodic boundary
condition in the $ z $-direction.  

\subsection{Uniform Flow}
\begin{figure}
\plottwo{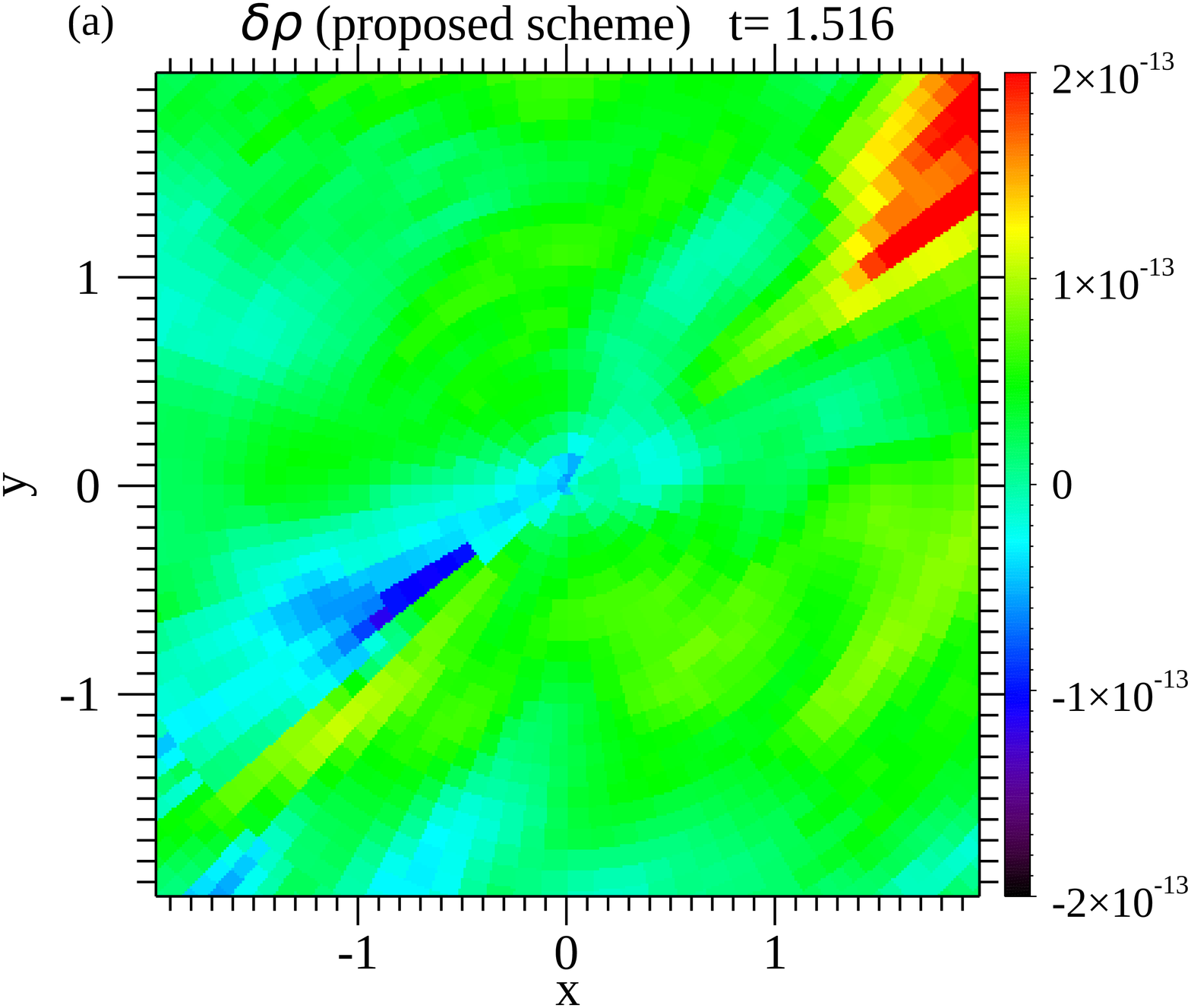}{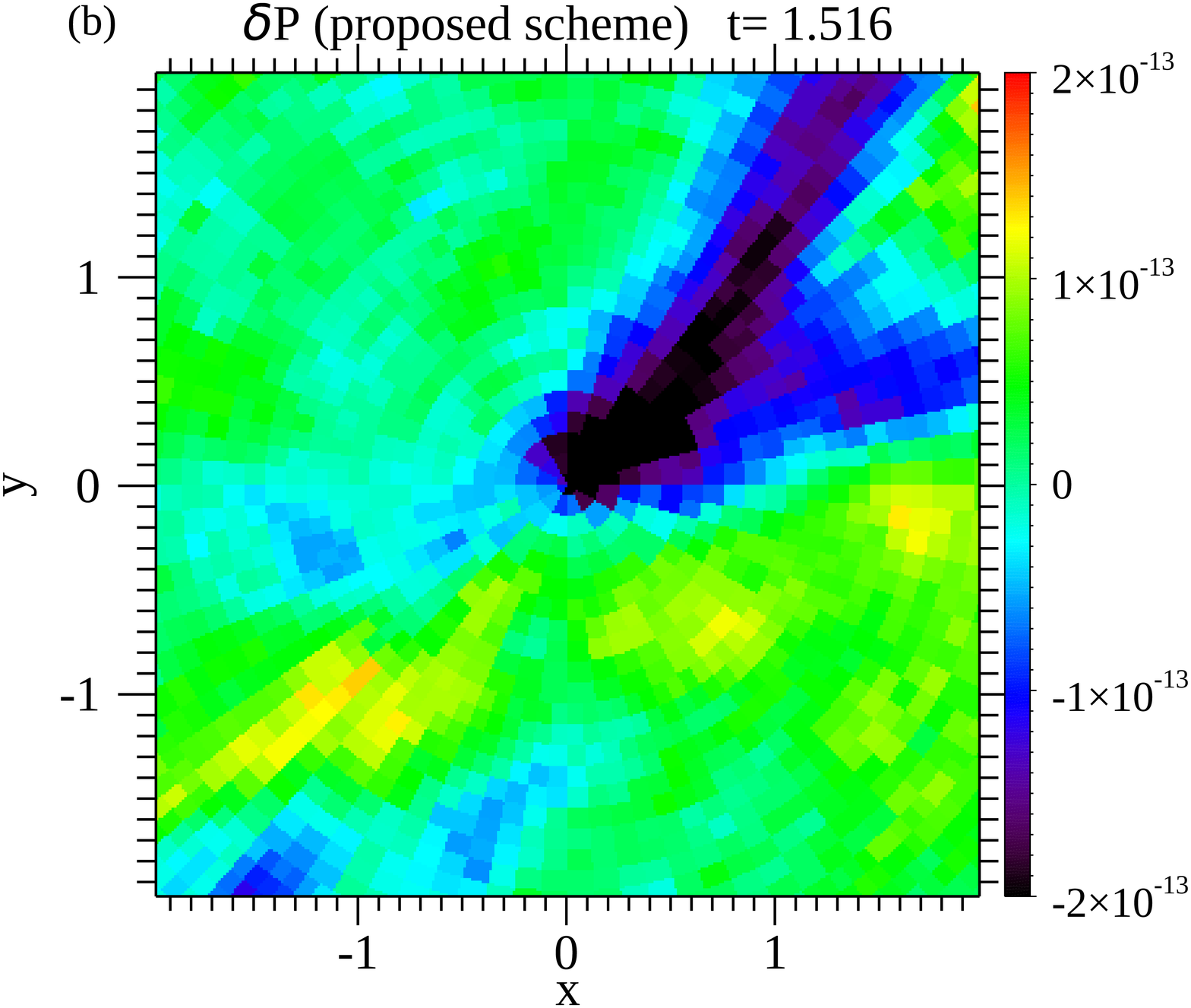}
\plottwo{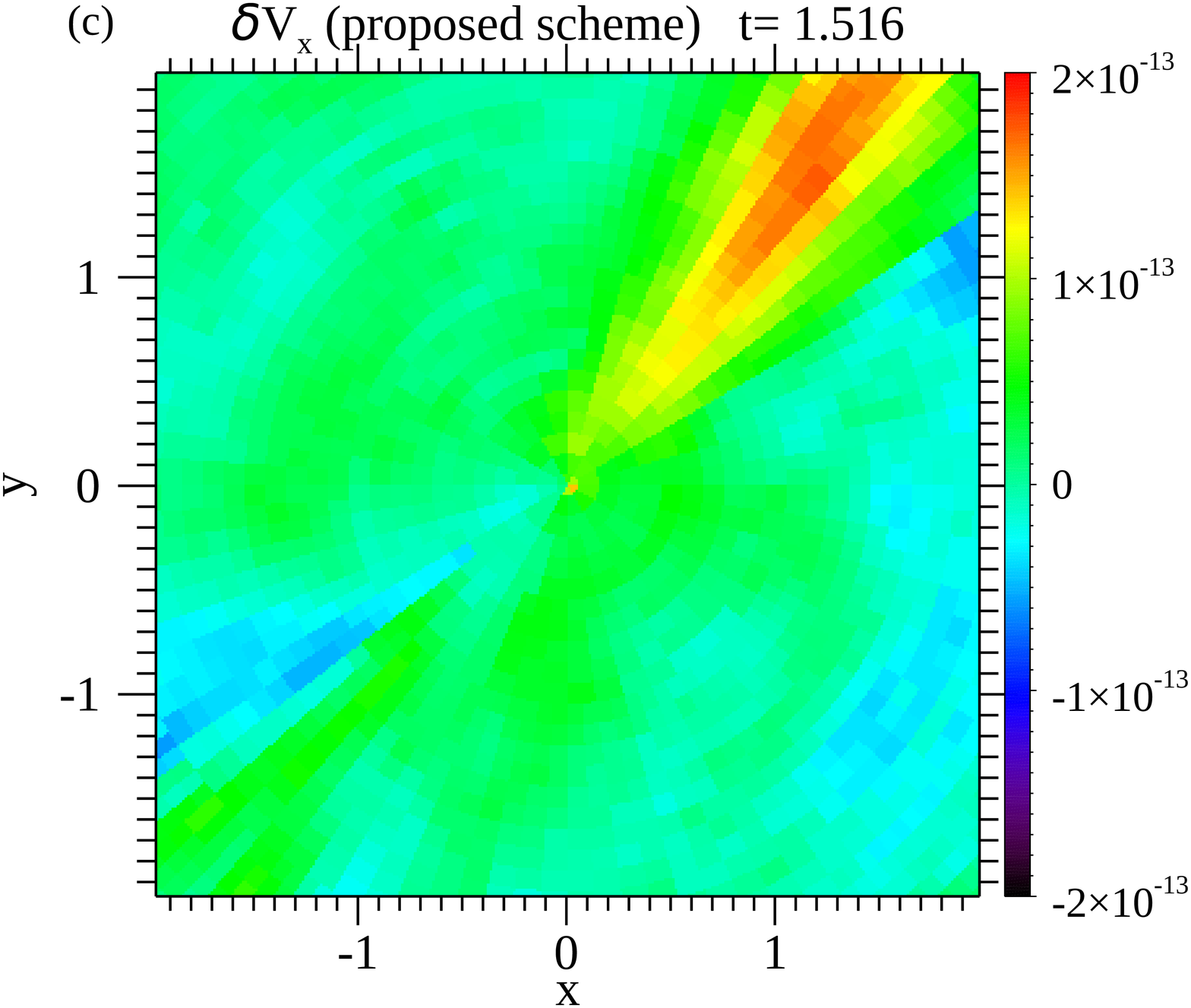}{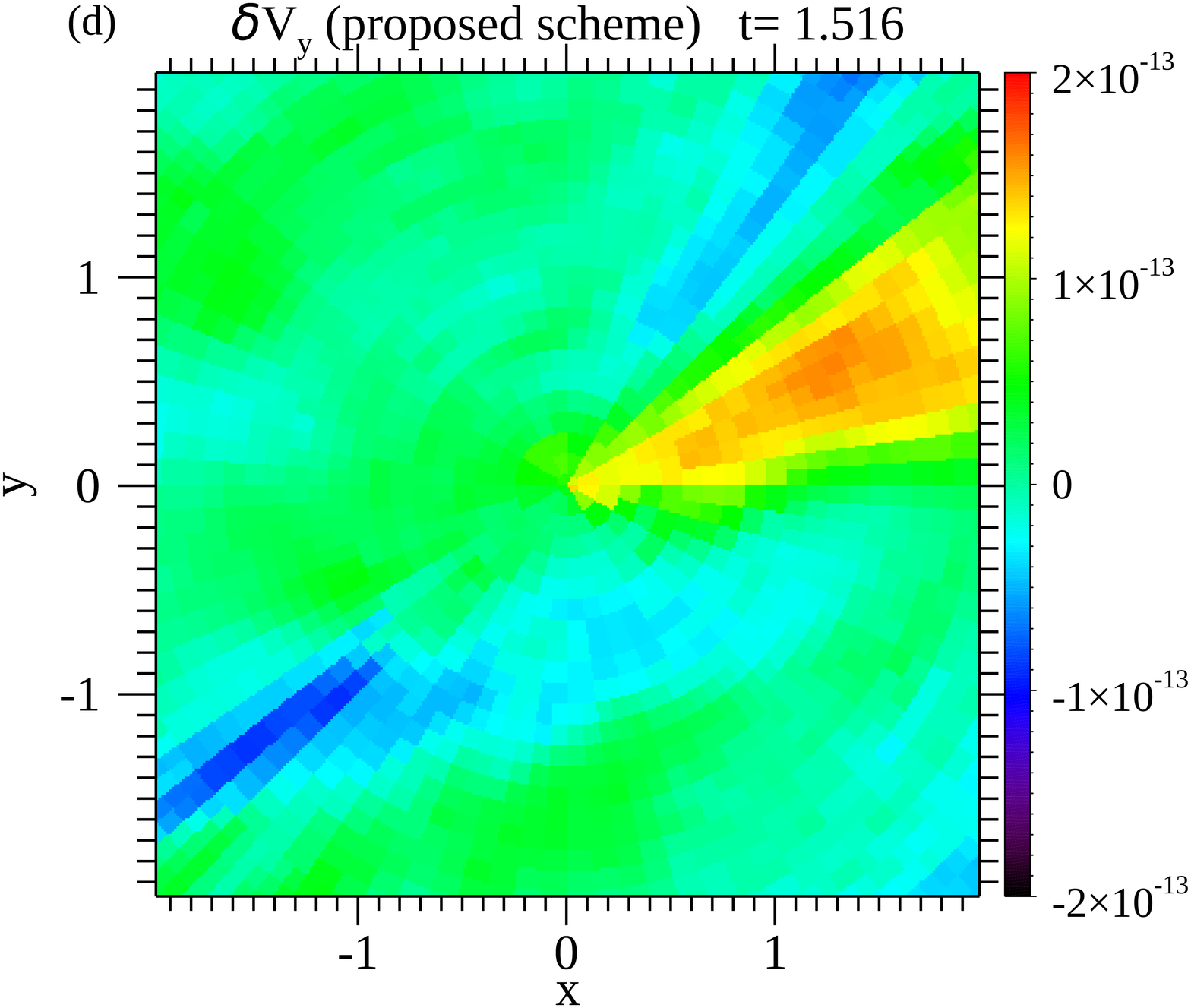}
\caption{Numerical error at the 1000th time step for uniform flow having 
$ \rho = 1 $, $ P = 1 $, and $ (v _x, v _y, v _z) = (4, 3, 2) $. 
Panels (a) and (b) denote the errors in $ \rho $ and $ P $, respectively,
while panels (c) and (d) denote those in $ v _x $ and $ v _y $, respectively. \label{uniformF}}
\end{figure}

First, we confirm that uniform flow is an exact solution of the discretized hydrodynamic
equations.   Figure \ref{uniformF} shows the numerical error in the example where 
the density and pressure are $ \rho = 1 $ and $ p  = 1 $, respectively, in the initial state.  
The initial velocity is set to be as follows,
\begin{eqnarray}
v _r & = & 4 \cos \varphi + 3 \sin \varphi , \\
v _\varphi & = & - 4 \sin \varphi + 3 \cos \varphi , \\
v _z & = & 2 ,
\end{eqnarray} 
so the velocity is uniform $ (v _x, v _y, v _z) = (4, 3, 2) $ in the Cartesian grid.
The numerical grid has 48 and 3 cells in the $ r $- and $ z $-directions, respectively.  The angular
resolution is $ \Delta \varphi = \pi / 3 $ in the innermost cells and $ \Delta \varphi = \pi / 96 $ in
the outer cells of $ i \ge 23 $.  The resolution in the intermediate range is set by Constraint (\ref{grid}).
The radial cell width in the innermost cell and the vertical cell width are set to be $ \Delta r _1 =
\Delta z = 0.1 $.  The CFL number is set $ {\rm CFL} = 0.3 $. 
Each panel of Figure \ref{uniformF} shows
the deviation from the analytic solution for $ \rho $, $ v _r $, $ v _\varphi $, and $ P $. The color
denotes the numerical error at $ t = 1.516 $ (at the 1000th time step).   They  are
$ | \Delta \rho | \le 1.141 \times 10 ^{-13} $, $ | \Delta P | \le 4.624 \times 10 ^{-13} $,
$ | \Delta v _r | \le 1.723 \times 10 ^{-13} $, and $ | \Delta v _\varphi | \le 1.608 \times 10 ^{-13} $ at the most. These numerical errors are caused by
the round-off error because we used double precision in our computation. The solution has no numerical error in $ v _z $. Thus, this example
shows that we have succeeded in free-stream preservation.

\begin{figure}
\plottwo{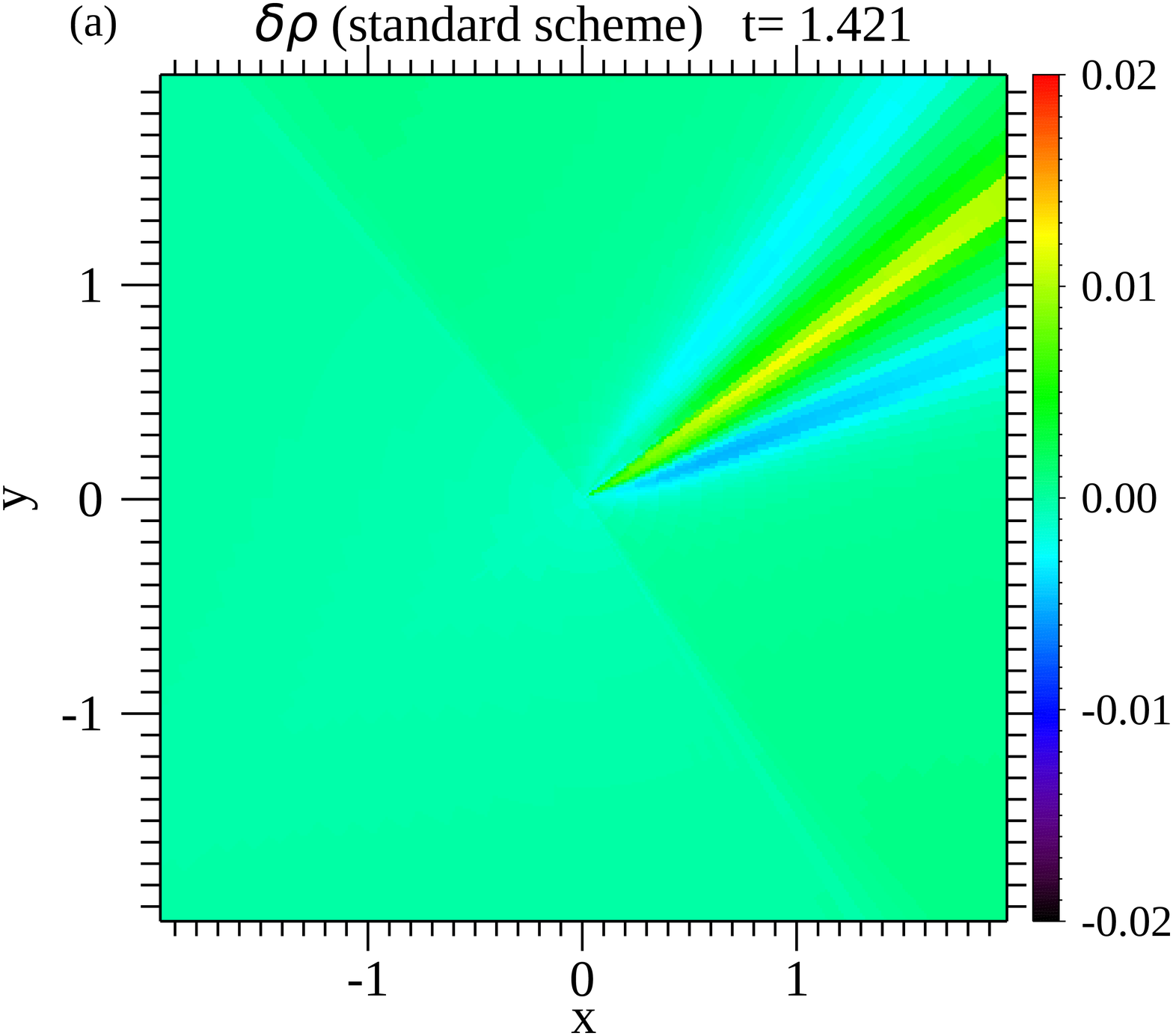}{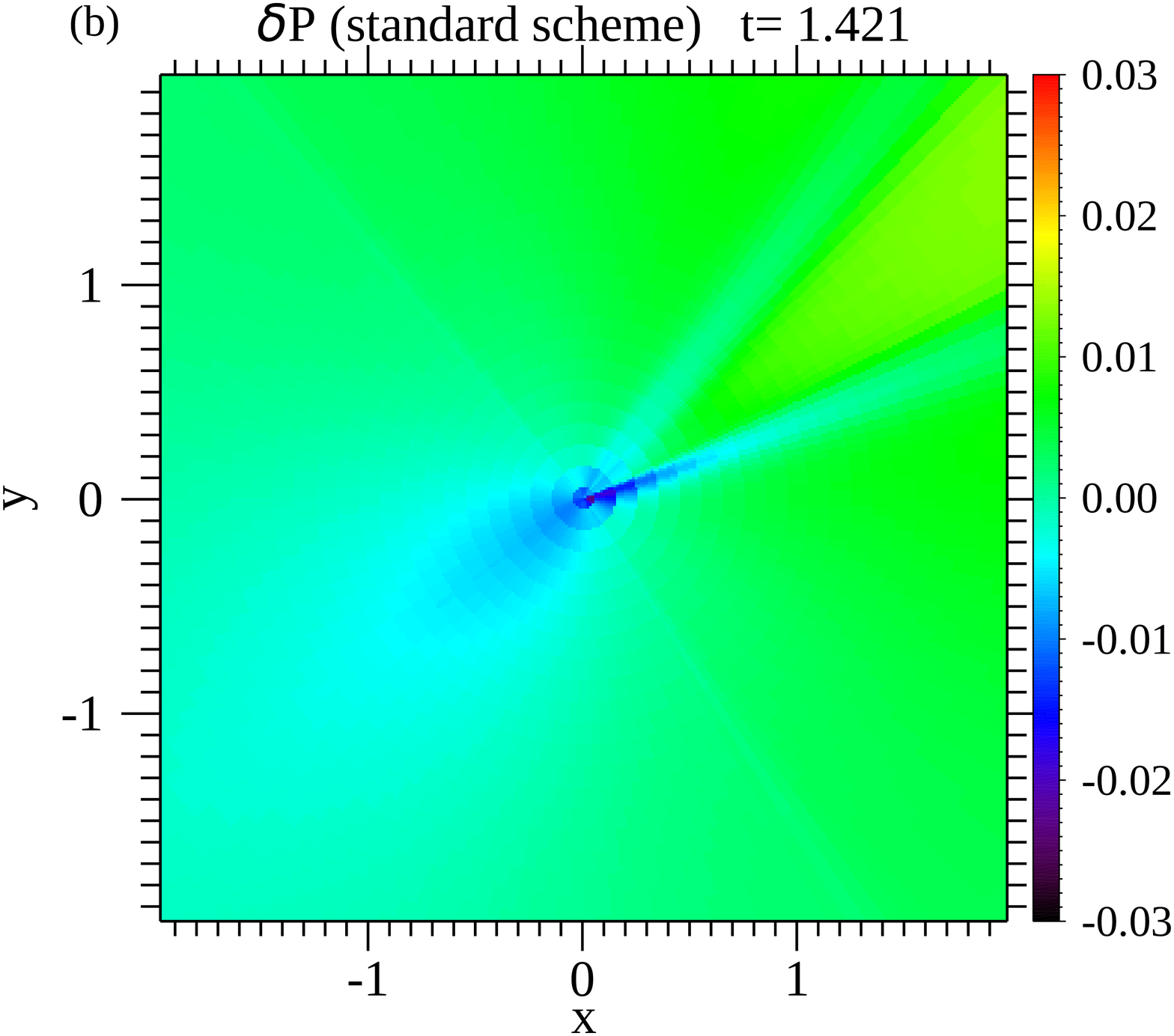}
\plottwo{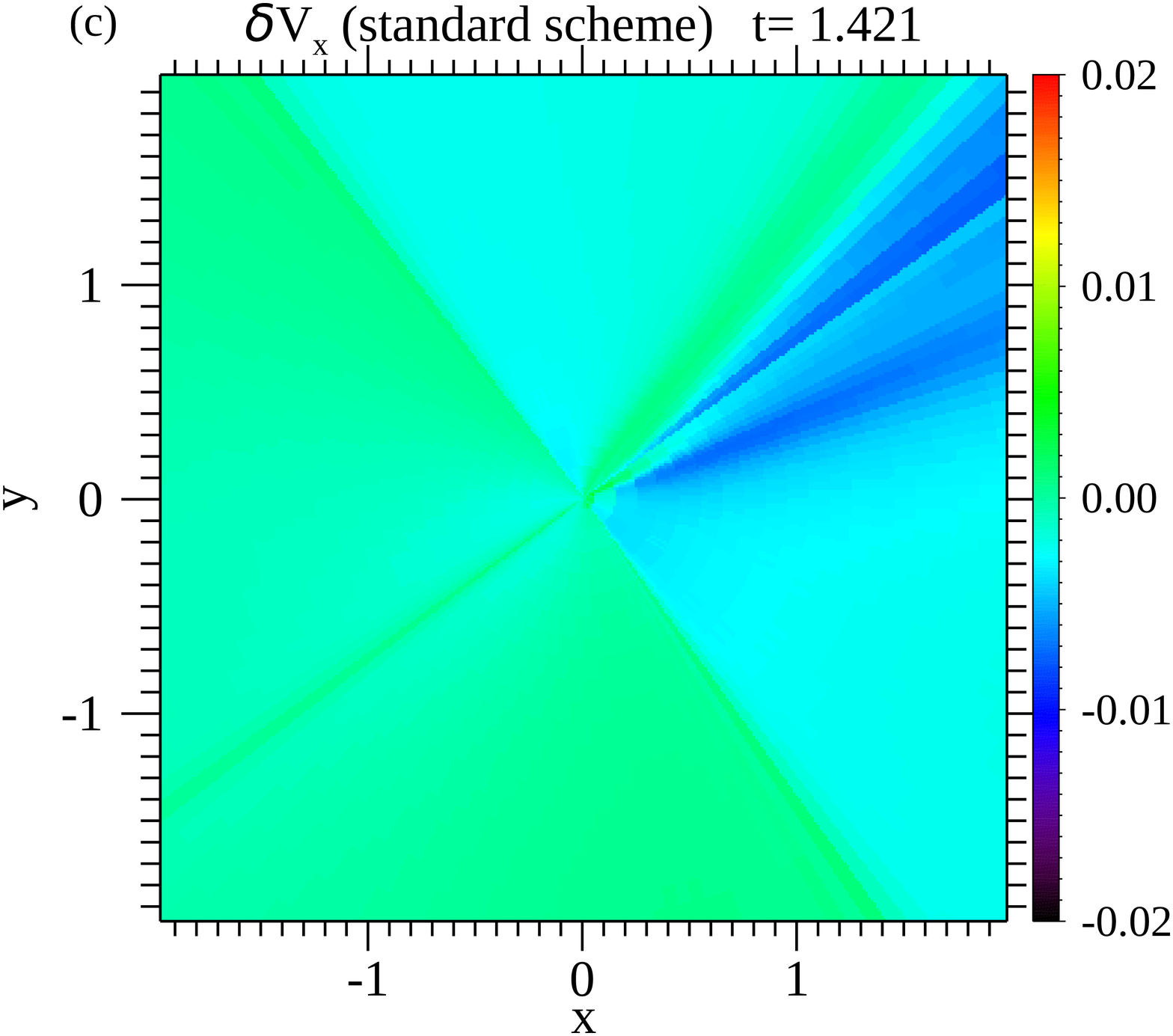}{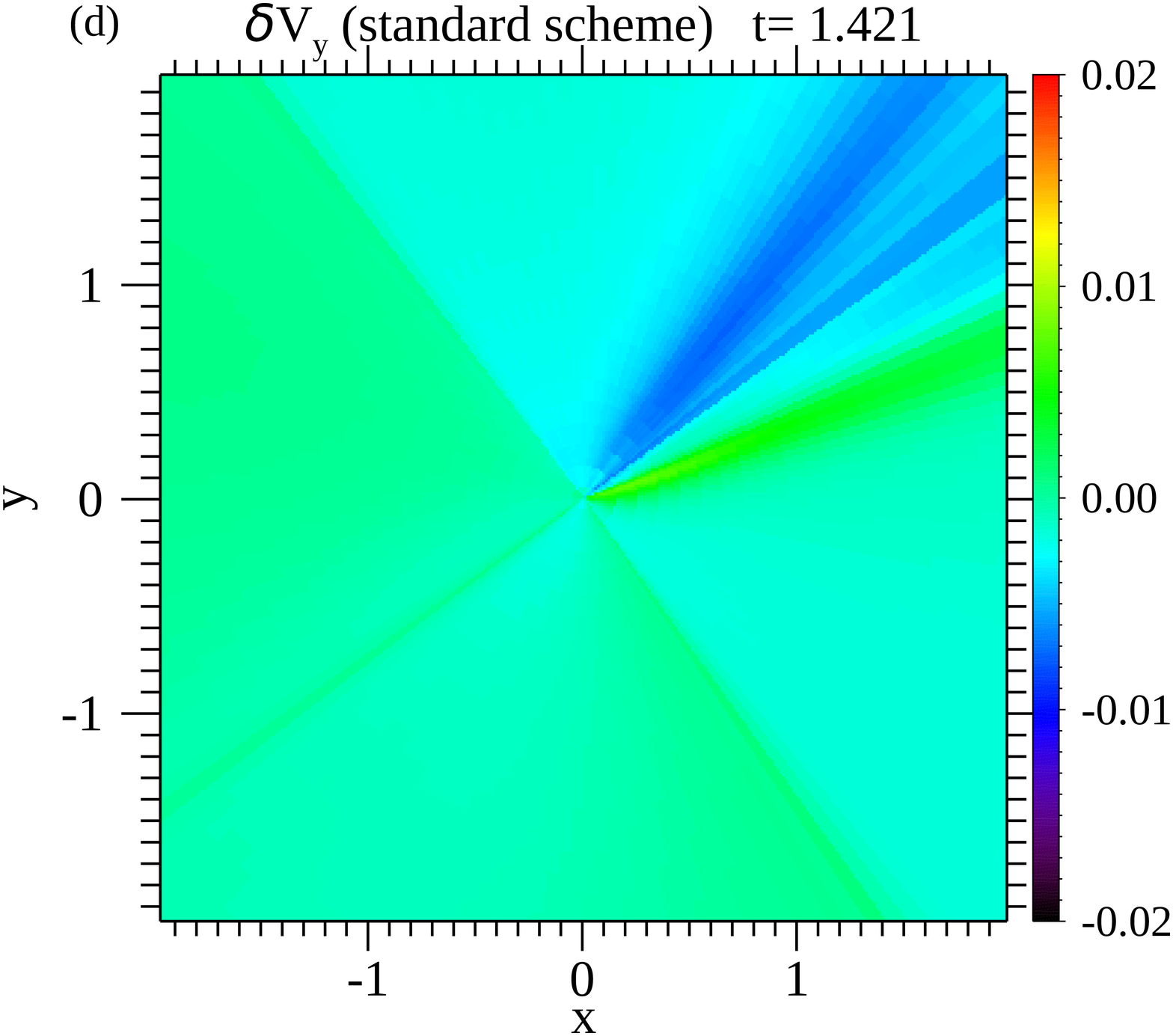}
\caption{The same as Figure \ref{uniformF} but for the solution obtained with
the current standard method.  The angular resolution is $ \Delta \varphi = \pi / 96 $.
\label{uniformFold}}
\end{figure}

We can solve the same problem with the current standard method if we take a high angular
resolution and are tolerant to small truncation errors.
Figure \ref{uniformFold} shows the numerical error of the solution at $ t = 1.421 $ 
given by the current standard
method with a high angular resolution, $\Delta \varphi = \pi / 96 $.  Despite this high angular
resolution, the density and pressure have relative errors of a few percent.   
Although the CFL number is again 0.3, the time step is short and the solution is obtained by 
$ 3.0 \times 10 ^4$ time steps.
Our new scheme gives a much more accurate solution at a much smaller computation cost.

We can solve this problem at a little lower angular resolution.  However, 
the truncation error increases roughly in proportion to $ (\Delta \varphi) ^{2} $. 
The pressure has the minimum value of $  P _{\rm min} = 0.884 $ and 0.519
for $ \Delta \varphi = \pi / 48 $ and $ \pi / 24 $, respectively, in this test problem.
When the Mach number of the flow is higher, the error is larger.  Thus, we need
still higher angular resolution.   Even when the flow is highly supersonic,
we can relax the CFL condition by reducing the angular resolution by employing
our new scheme.   Although the complexity of our new scheme 
slightly increases the computational cost per time step, we can greatly benefit from the decrease in the number of time steps.

The correction factors have substantial effects in the improvement
described above.  When they are suppressed artificially, the proposed 
scheme fails at the fourth time step. The failure is due to the
supersonic flow.  We need to include the correction factors when
the flow is highly supersonic.

\subsection{Sod Shock Tube Problem}

Next, we examine our new scheme using the Sod shock tube problem initialized by the following condition,
\begin{eqnarray}
\left( \rho, v _r, v _\varphi, v _z, P \right) & = &
\left\{ 
\begin{array}{lc}
(0.1,~0.0,~0.0,~0.0,~0.125) & (0 \le \varphi < \pi) \\
(1.0,~0.0,~0.0,~0.0,~1.0) & (\pi \le \varphi < 2 \pi) \\
\end{array}
\right. 
\end{eqnarray}  

\begin{figure}
\plottwo{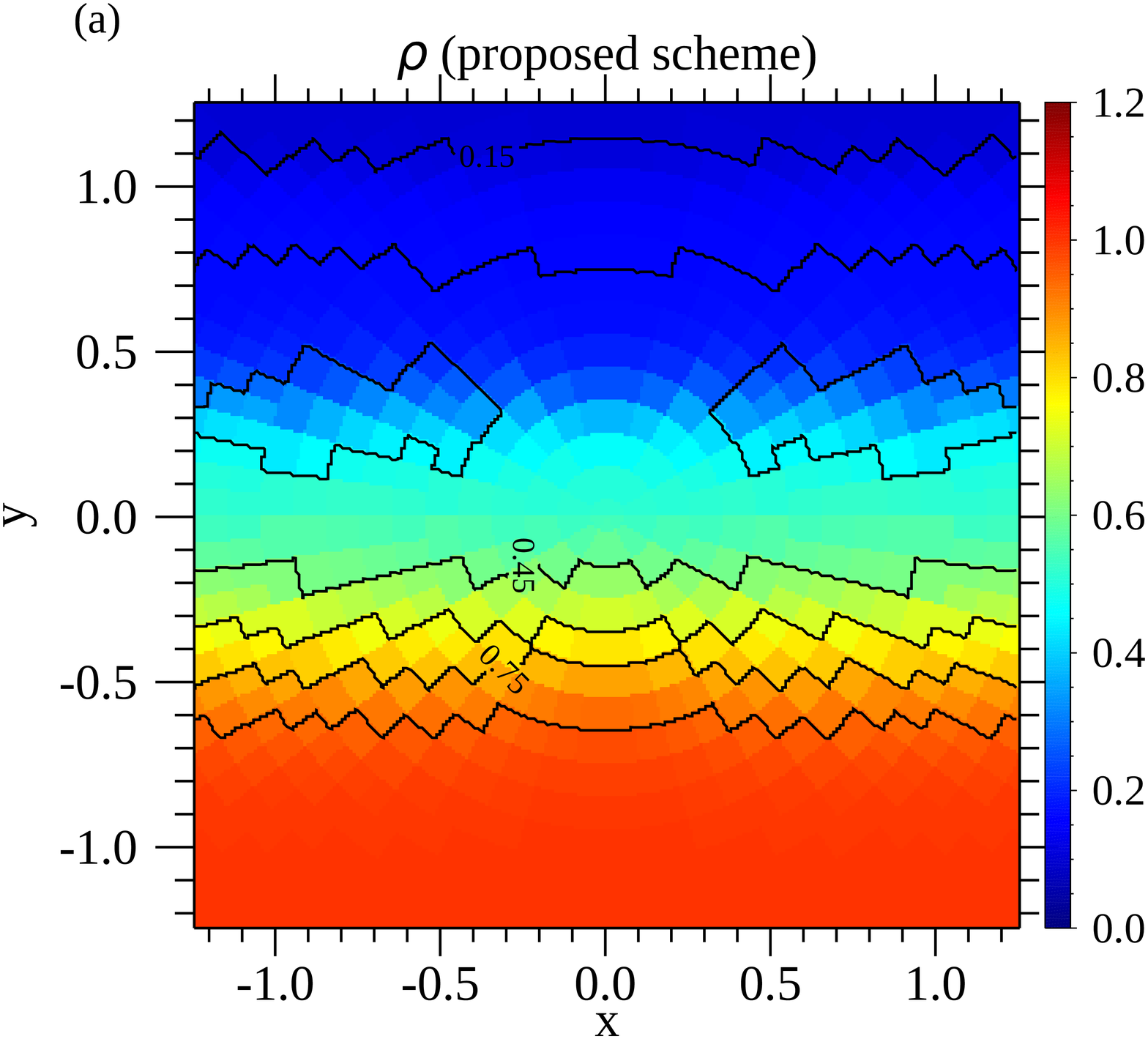}{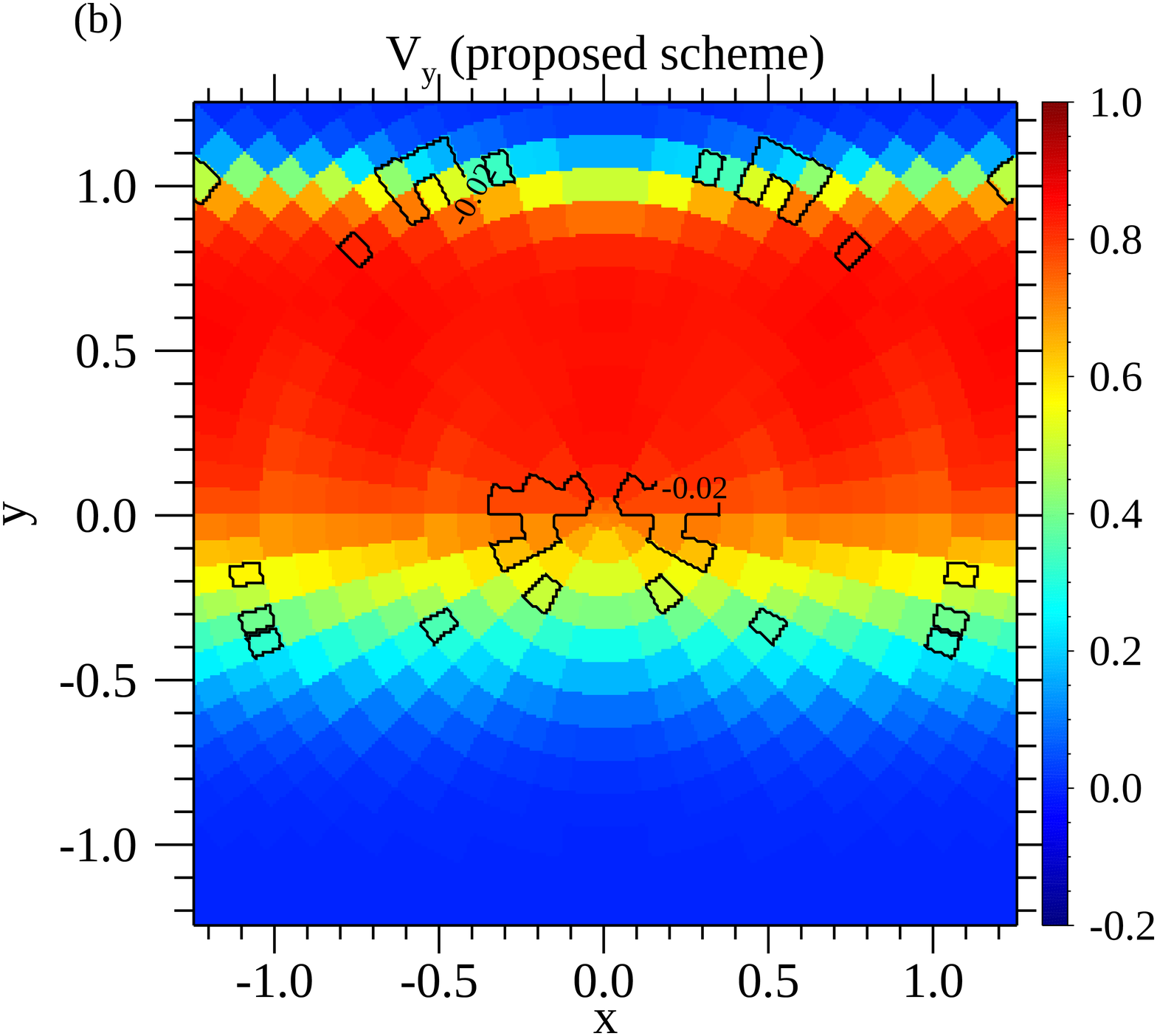}
\caption{The solution of the Sod shock tube problem obtained by the proposed scheme.  
The color and contours show $ \rho $ and $ P $ at $ t = 0.484 $, respectively, in panel (a), while
they show $ v _y $ and $ v _x $, respectively, in panel (b). \label{shock1}}
\end{figure}

Figure \ref{shock1} shows the solution given by our new scheme at $ t =  0.484 $. The CFL number is taken to be 0.5 because this is a 2D problem. The left panel shows
the density by color and the pressure by contours.  The right panel does $ v _y $ by color
and $ v _x $ by contours.   
The contours of $ \rho $, $ P  $, and $ v _y $ are nearly straight,
although zigzagged along the cell boundary.   The velocity tangential to the wave front is smaller
than $ | v _x | < 0.037 $.    The highest angular resolution is $ \Delta \varphi = \pi /48 $ in the
region of $ r > 2.3 $.

\begin{figure}
\plottwo{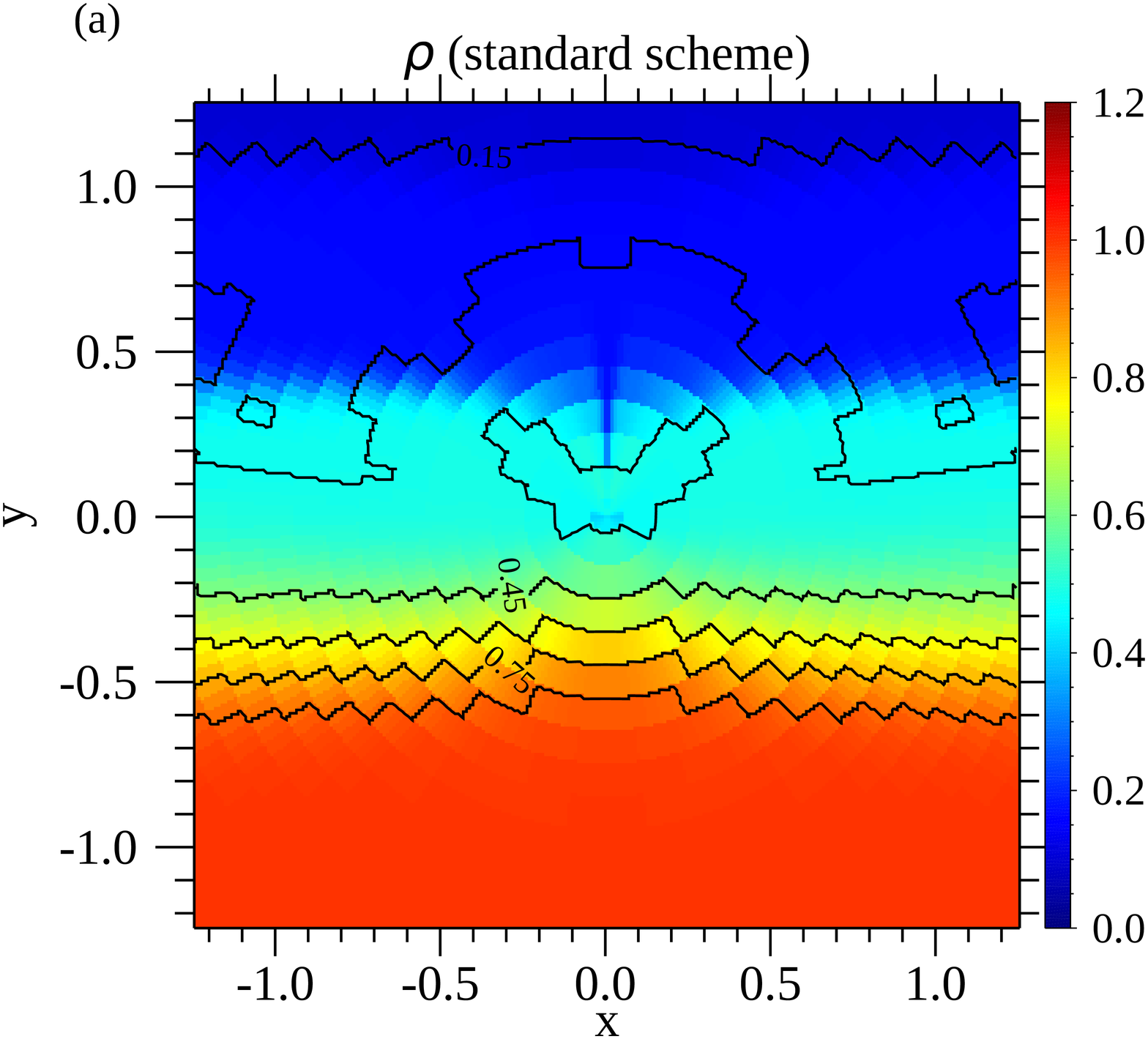}{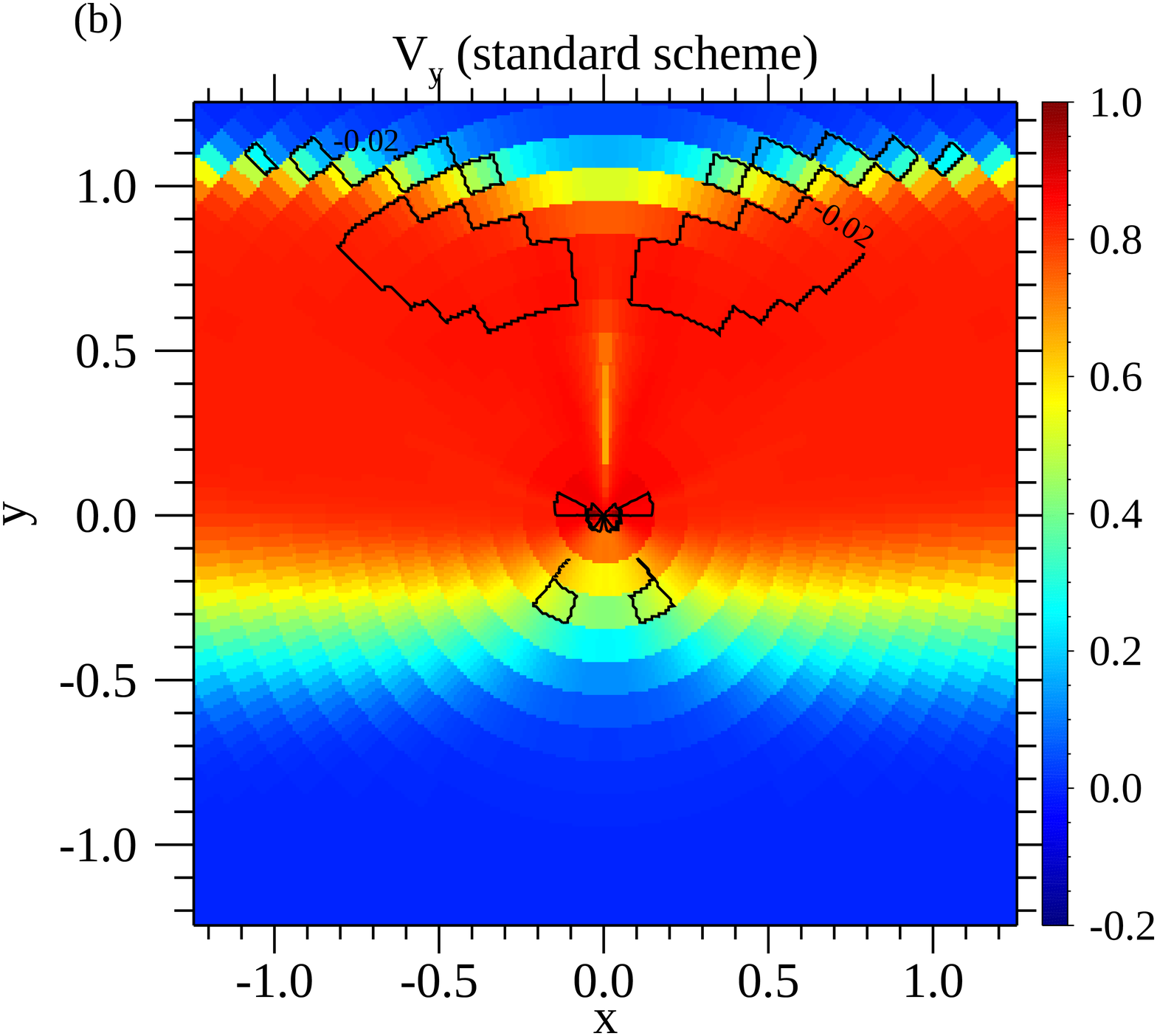}
\caption{The same as Figure \ref{shock1} but for the solution obtained by the current standard method at $ t = 0.503 $.
\label{shock2}}
\end{figure}

We compare this solution with that obtained by the current standard method shown in 
Figure \ref{shock2}.   The angular resolution is uniform at $ \Delta \varphi = \pi / 96 $ in
the whole region, and the CFL number is set to be 0.5.  
Figure \ref{shock2} shows $ \rho $, $ P $, $ v _y $, and $ v _x $ at
$ t = 0.503 $.  Despite the lower angular resolution, our new scheme provides a better feature.
The current standard method leaves a spike in the direction of $ \varphi = \pi / 2 $ when
the resolution is high.  This spike is because of the anisotropy of the numerical cell.  Note that
the numerical cells are highly elongated in the $ r $-direction.  
We think that the spike 
remains even in the limit of $ \Delta \varphi \rightarrow 0 $.  In other words, we are
afraid that the current standard method does not converge to the exact solution.

\begin{figure}
\plottwo{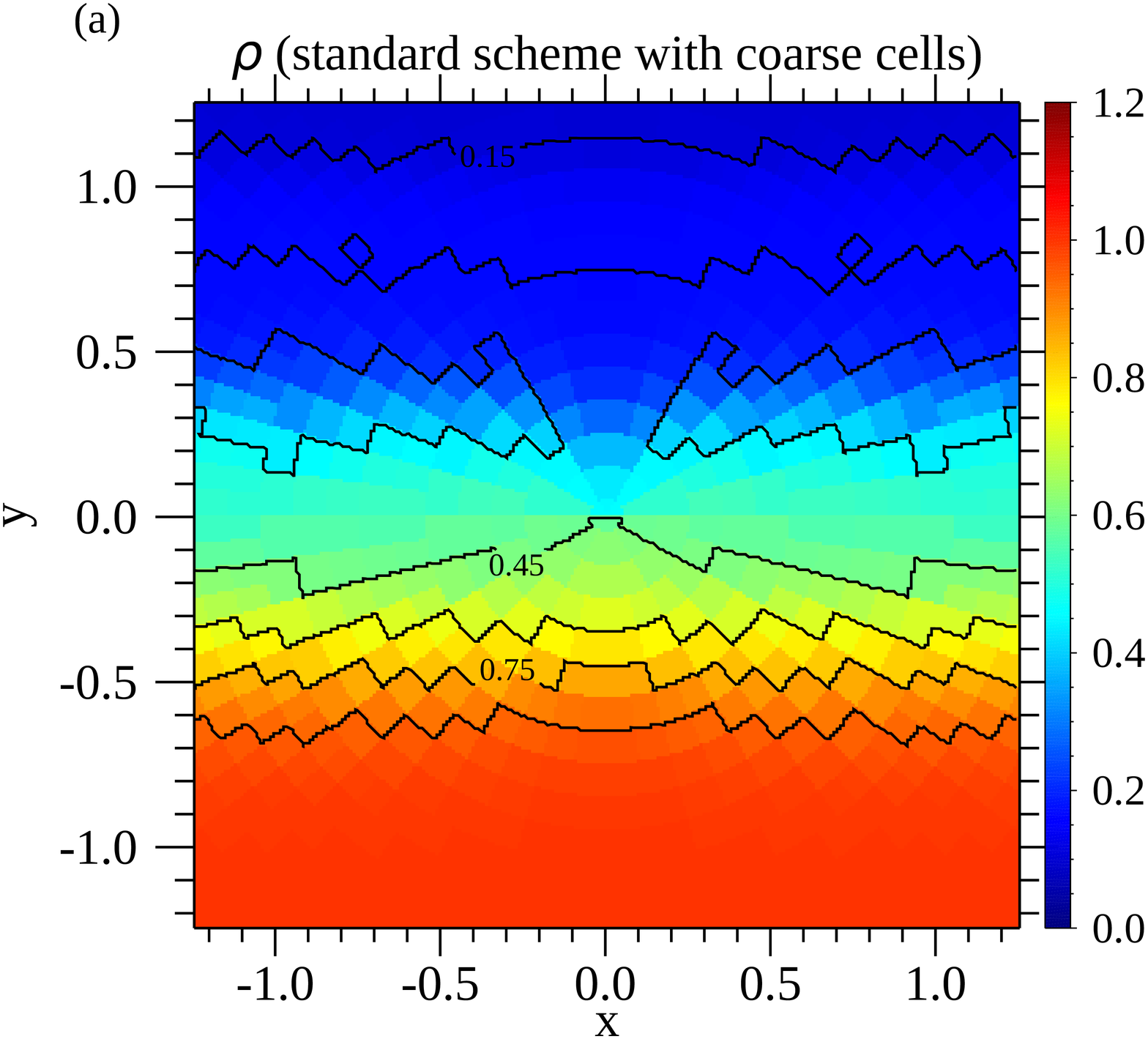}{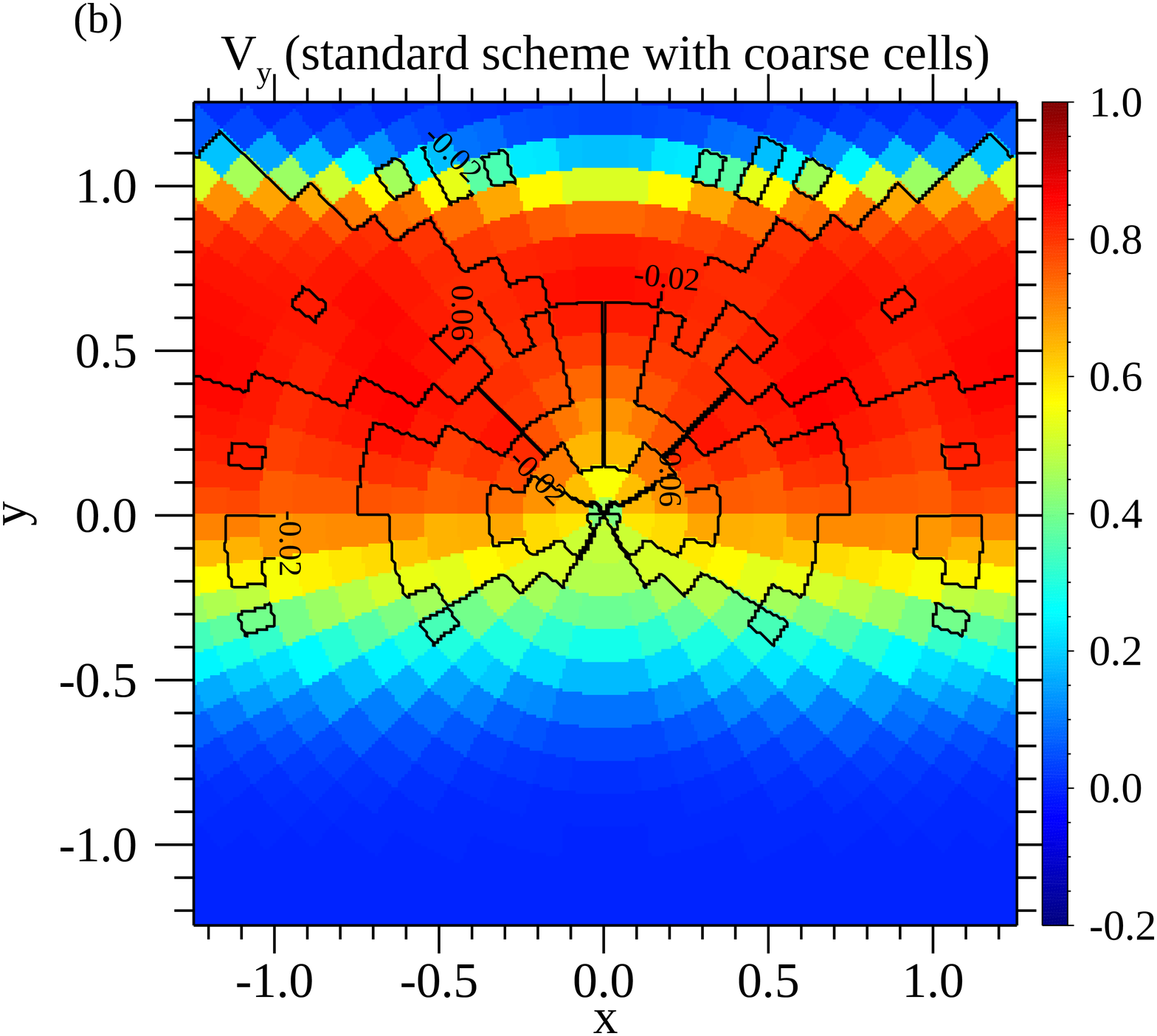}
\caption{The same as Figure \ref{shock1} but for the solution obtained with the current standard method
on the AMR-type grid.
The color and contours denote $ \rho $, $ P $, $ v _x $, and $ v _y $ at $ t = 0.469 $.
\label{shock3}}
\end{figure}

This spike feature is weakened if we use the AMR-type grid shown in Figure \ref{AMR}. 
Figure \ref{shock3} shows the solution obtained with the current standard on the
AMR-type grid.  The color and contours denote $ \rho $, $ P  $, $ v _y $, and $ v _x $
at $ t = 0.469 $. The pressure distribution does not show such a spike, but the isobaric contour is largely
curved.

\begin{figure}
\plottwo{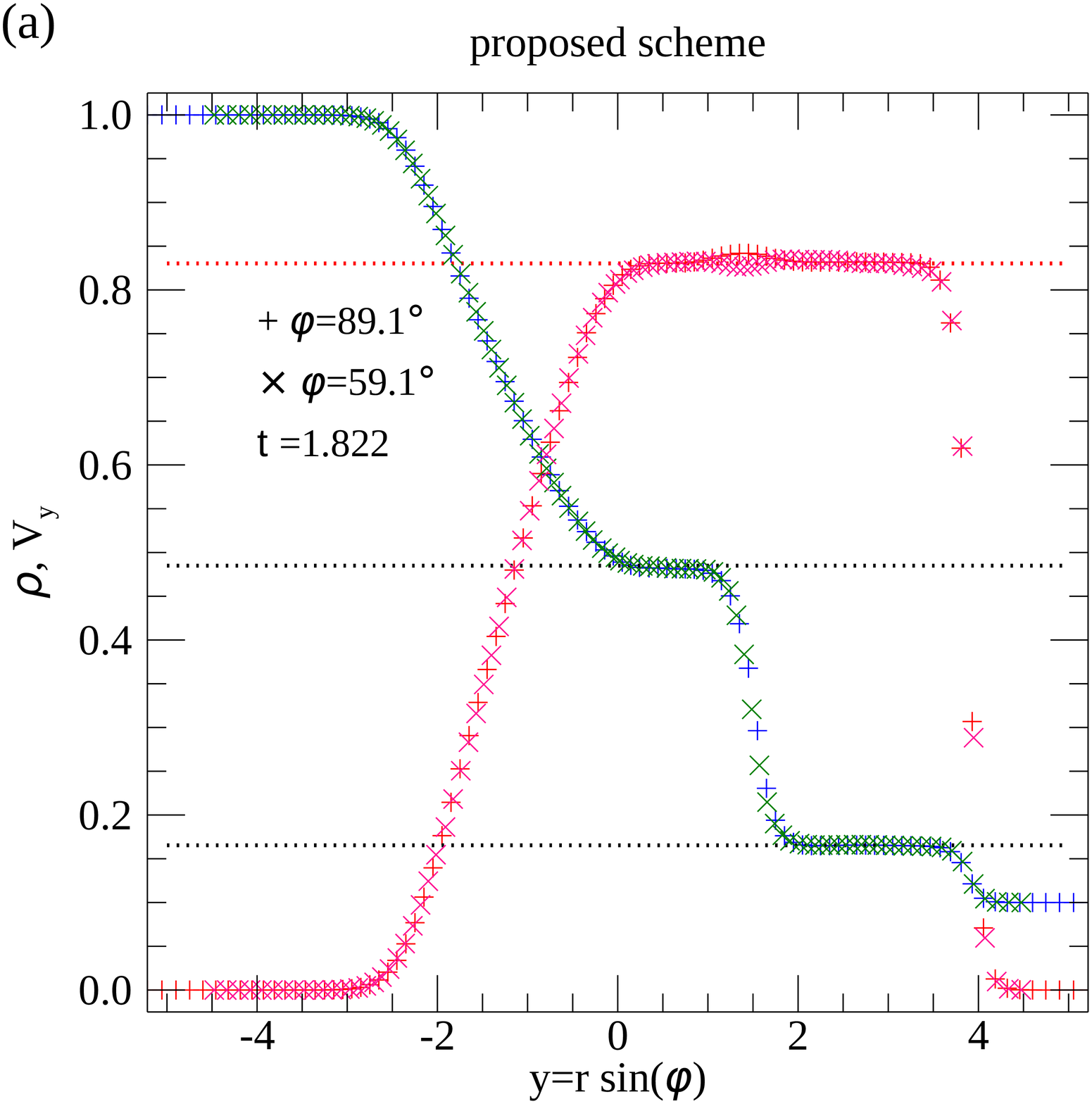}{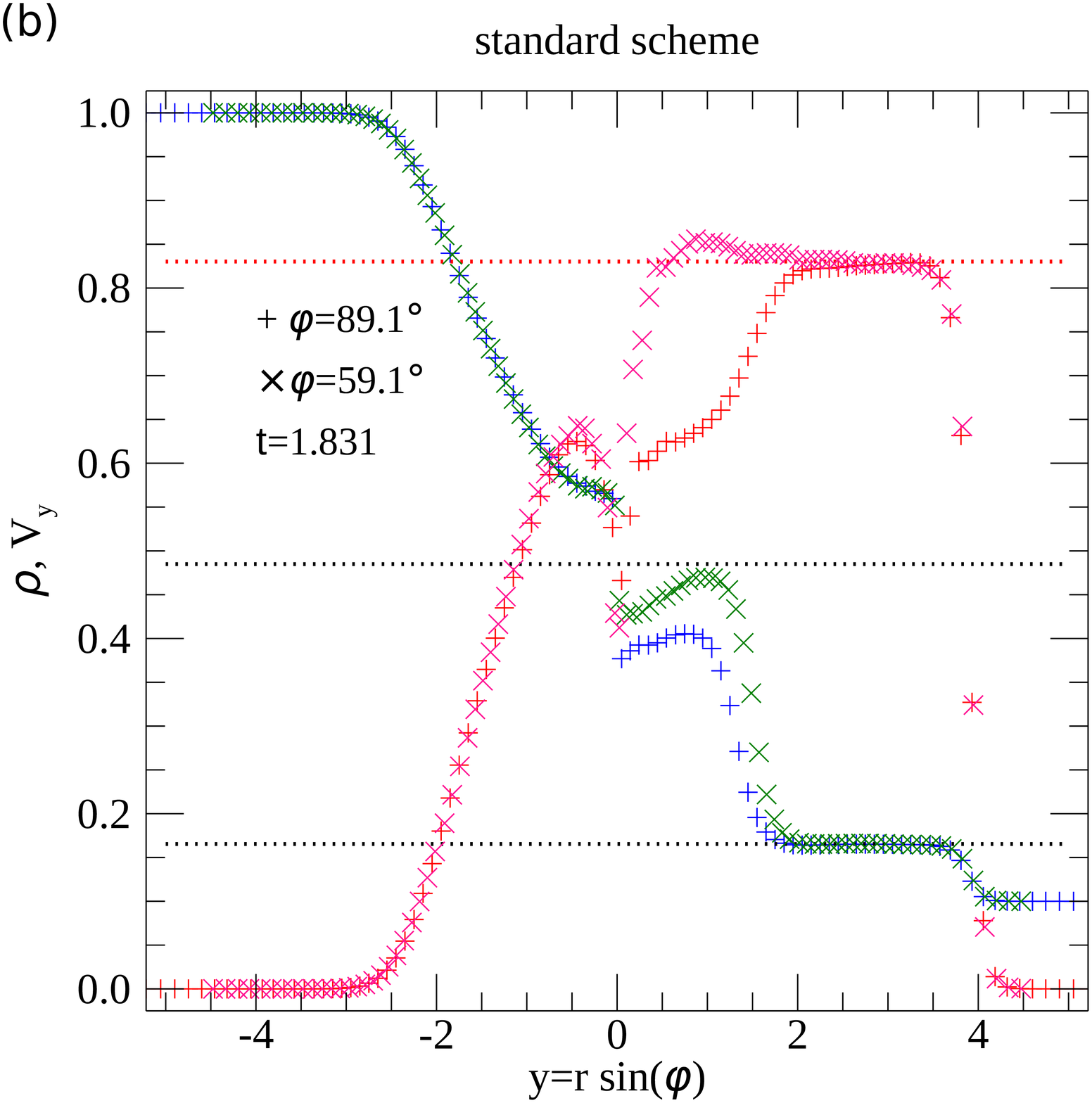}
\caption{The solution to the shock tube problem.  Panel (a) shows the solution obtained with 
our new scheme, while panel (b) shows that obtained with the current standard method on the
AMR-type grid. The crosses and plus signs denote the values on the lines of 
$\varphi = 89.^\circ 1 $ and $59.^\circ 1$, respectively, as a function of
$ y = r \sin \varphi $. The green and blue symbols denote the density,
while the red and orange do $ v _y $. \label{shock4}}
\end{figure}

We compare the solutions shown in Figures \ref{shock1} and  \ref{shock3} more quantitatively
at a later stage.   Figure \ref{shock4} shows the density and velocity on the lines of $\varphi = 59.^\circ 1$ and $89.^\circ 1$ by crosses and plus signs, respectively. The density is denoted by blue and green, while the velocity by red and orange. The left
panel shows the solution obtained with our new scheme, while the right panel shows that obtained
with the current standard method.  The red dotted lines denote the velocity of the contact
surface, while the black {dotted} lines denote the density forward and backward of the contact
surface. It is clear that our new scheme provides a better approximation to the exact solution.
The current standard method gives a poor solution around the origin. 
{Even at $ t = 4.499$, the density ranges from 0.365 to 0.547 within the innermost cells around the $ z $-axis.  
The density ranges from 0.467 to 0.527 around the $z $-axis depending on $\varphi $ at $ t = 1.677 $ in the solution obtained with the uniform grid with {$ \Delta \varphi = \pi/96$} (cf. Figure \ref{shock2}). 
The proposed scheme provides a solution converging to the same density in the innermost
cells.  The solution converges
to the exact one until the wave reflected from the outer boundary reaches.  The L$_1$ norm of the error
reduces approximately in proportion to the cell width.  The error has an appreciable
amplitude around a narrow region around the contact discontinuity and shock front.
Although the error is due to the jump in the exact solution and inevitable, 
the region of a substantial error is limited in the shock tube problem in 1D.    

We can confirm the superiority of the proposed scheme over the current standard scheme 
also in the tangential velocity.
The tangential velocity is smaller than $ \left| v _x \right| \leq 3.335 \times 10 ^{-3} $
in the inner region of $ r < 0.2 $ at $ t = 2.262 $ in the solution obtained with the
proposed scheme.  However, it has the maximum value, $ 1.196 \times 10 ^{-1} $ and $ 4.303 \times 10 ^{-2} $, 
in the same region in the solution obtained with the current standard scheme on the
AMR-type grid at $ t = 2.273 $ and that on the uniform angular resolution of $ \Delta \varphi / 96 $
at $ t = 2.230 $, respectively.  We cannot eliminate this spurious tangential velocity perfectly, since 
the wave front is inclined with respect to the cylindrical grid.  The spurious
tangential velocity appears only temporally in the solution obtained with the proposed
scheme, while it remains persistently around the $ z $-axis in those obtained
with the current standard scheme.}

\subsection{Expanding Outflows}

Next, we apply our new scheme to a rotating gas sphere expanding into a conical
cavity, the axis of which is inclined by $ \chi $ from the $ z $-axis.  The initial
density and pressure are set to be 
\begin{eqnarray}
\left( \rho, P \right) & = & \left\{
\begin{array}{lll}
(\rho _{\rm m},  P _{\rm h}) & \mbox{for} & \sqrt{r ^2 + z ^2} < a   \\
(\rho _{\rm h}, P _{\rm \ell}) &\mbox{for} &  \sqrt{r ^2 + z ^2} \ge a \, \mbox{and} \,  | \cos \theta | \le \cos \alpha \\
(\rho _{\rm \ell}, P _{\rm \ell}) & \mbox{for} &  \sqrt{r ^2 + z ^2 } \ge a \ \mbox{and}  \, | \cos \theta | > \cos \alpha
\end{array}
\right. , \end{eqnarray}
\begin{eqnarray}
\cos \theta & = & \frac{r \sin \chi \cos \left( \varphi - \lambda \right) + z \cos \chi}{\sqrt{r ^2 + z ^2}} , 
\end{eqnarray}
{where $ \alpha $ denotes the
half opening angle and set to be $ \alpha = \pi/ 4$.  The axis of the cavity is inclined by $ \chi $ from the $z$-axis in the direction of $ \varphi = \lambda $.}
The initial velocity is set to vanish in the outer region of $ r ^2 + z ^2 \ge a $.  The central
dense gas is assumed to rotate with angular velocity, $ \omega $, around the axis of the cavity and
expands homogeneously with the rate, $ \nu $, 
\begin{eqnarray}
\bmv{v} & = & \nu \left( r \bmv{e} _r + z \bmv{e} _z \right) +
\omega  \left\{ z \sin \chi \sin (\varphi - \lambda)  \, \bmv{e} _r 
+ [z \sin \chi \cos (\varphi - \lambda ) + 
r \cos \chi ] \bmv{e} _\varphi 
- r \sin \chi \sin ( \varphi - \lambda ) \, \bmv{e} _z \right\} ,
\end{eqnarray}
in the inner region of $ \sqrt{r ^2 + z ^2 } \le a $ at the initial stage.  Thus, the velocity distribution is 
also symmetric around the axis of the conical cavity.  The density distribution is
given by $ \left( \rho _{\rm \ell}, \rho _{\rm m}, \rho _{\rm h} \right)$ =
$ (0.1,~0.2,~1.0)$ while the
pressure distribution is given by $ (P _{\rm \ell} , P _{\rm h} ) = $ 
= $ (0.2,~0.5 )$. The half opening
angle of the cavity is set to be $ \alpha = \pi / 4$.   The expansion and rotation are characterized by
$ \nu = \omega = 5 $. The initial radius of the sphere is set to be $ a = 1 $. Note that the
initial sound speed is 2.041 in the gas sphere, while both the expansion
and rotation velocity are 5. 

\begin{figure}
\plottwo{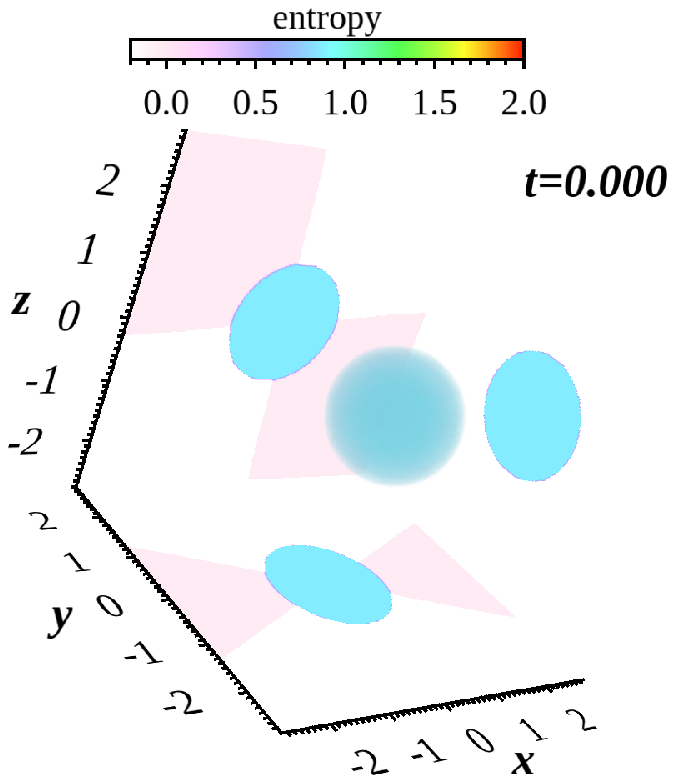}{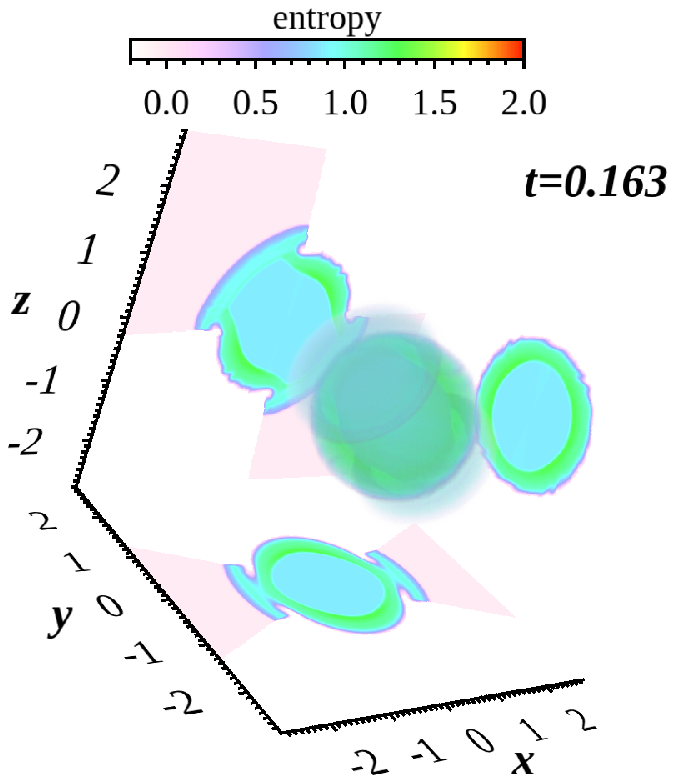}
\plottwo{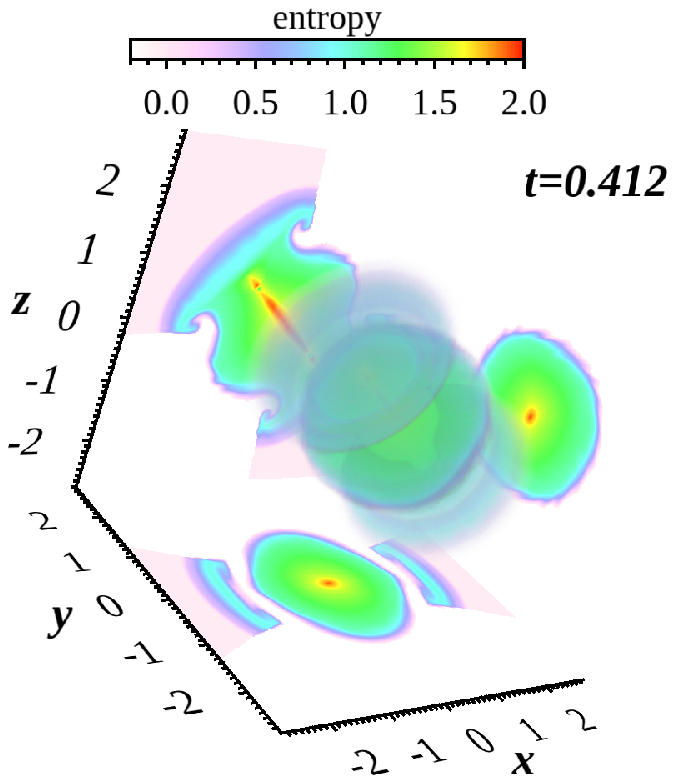}{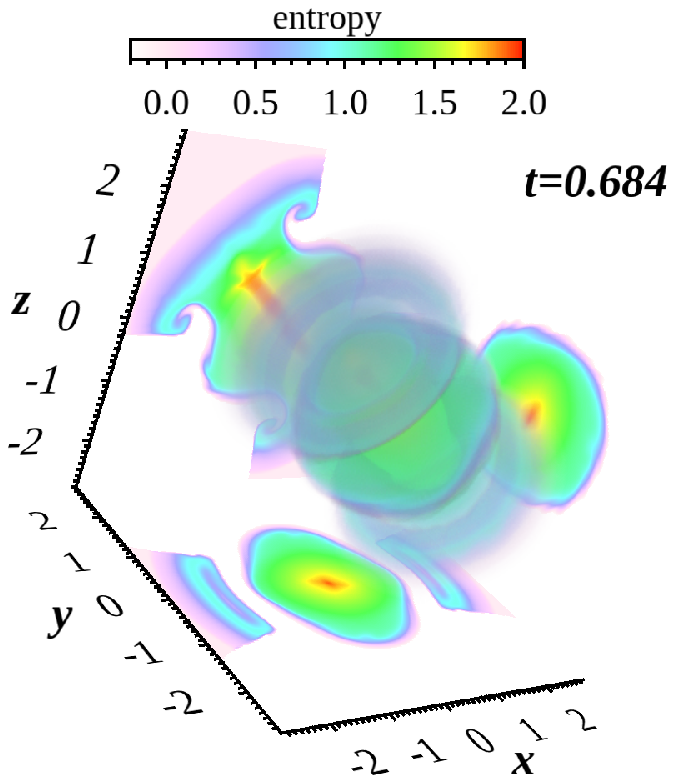}
\caption{The entropy distribution is shown for the expanding outflow
by a series of snapshots.  Each panel shows the entropy by the volume rendering and cross
sections. \label{outflow1}}
\end{figure}
Figure \ref{outflow1} shows the solution obtained with the proposed scheme by a series of snapshots.
The conical cavity is set as $ (\chi, \lambda)  = [0.9, (1+1/384)\pi] $.  The innermost numerical cell has
the width, $ \Delta r = \Delta z  = 0.025 $ and $ \Delta \varphi = \pi / 3 $.  The angular resolution 
is $ \pi / 192 $ in the region $ r >  1.138 $. The CFL number is set to be 0.3.  The color denotes the entropy, $ s \equiv \log P - \gamma \log \rho $.
The central object in each panel shows the entropy distribution by the volume rendering, while the side objects 
show the entropy distribution by the cross sections on the $ x = 0 $, $ y =0 $, and $ z = 0 $ planes.
The initial gas sphere expands into the conical cavity while spinning around the axis.
Our new scheme follows this expansion smoothly even at the resolution of $ \Delta \varphi = \pi /3 $.

\begin{figure}
\epsscale{0.85}
\plottwo{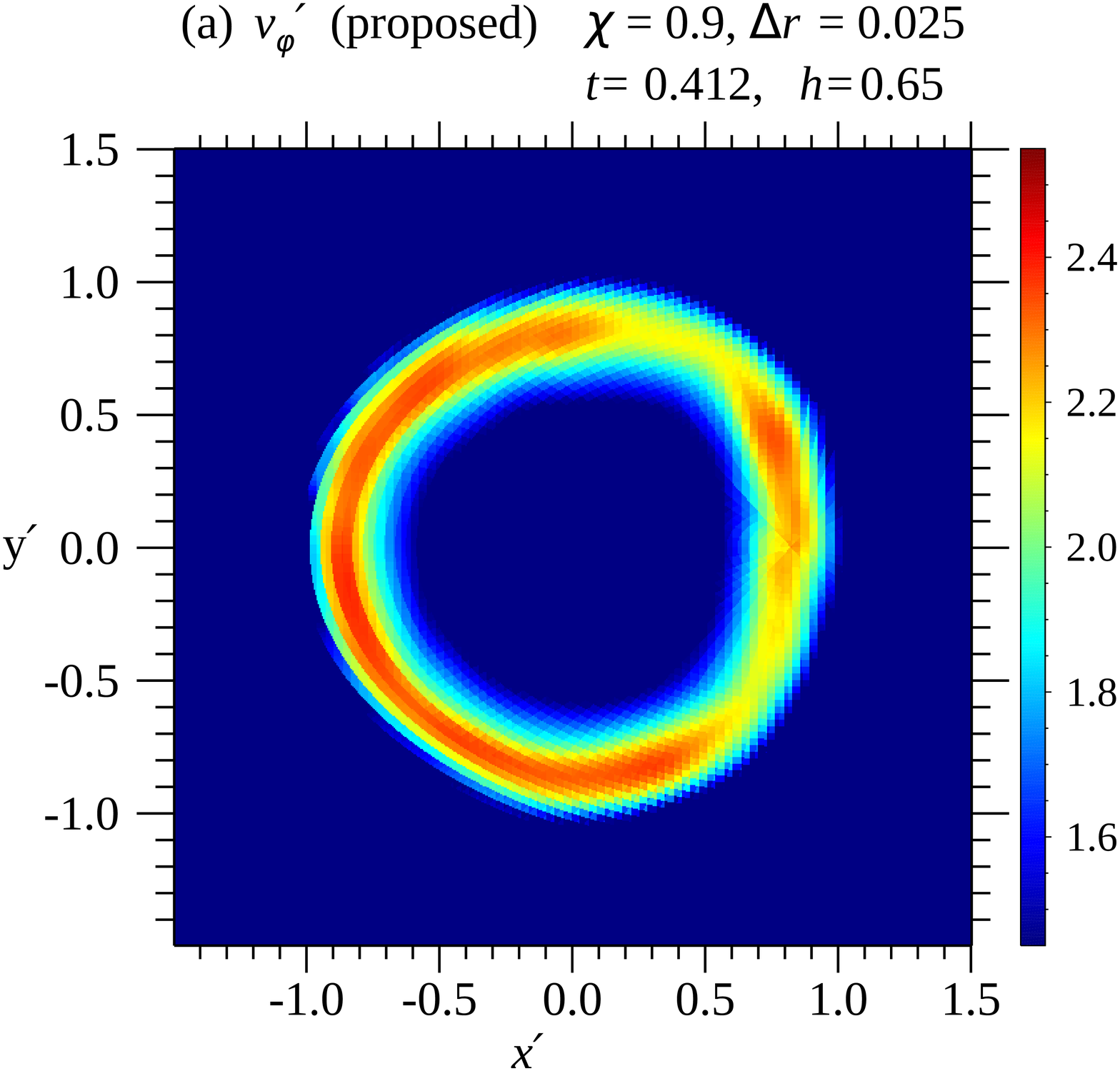}{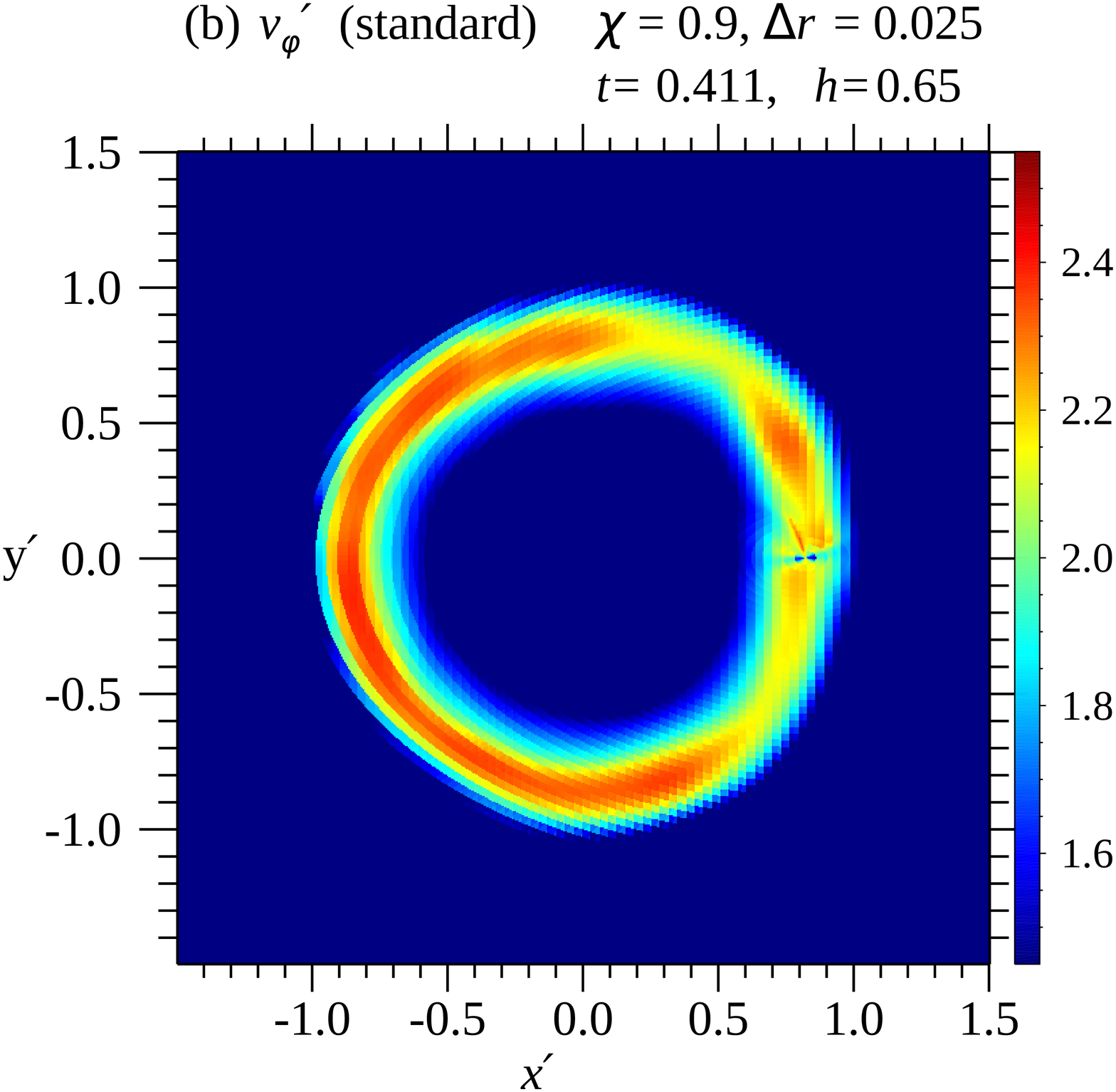}
\plottwo{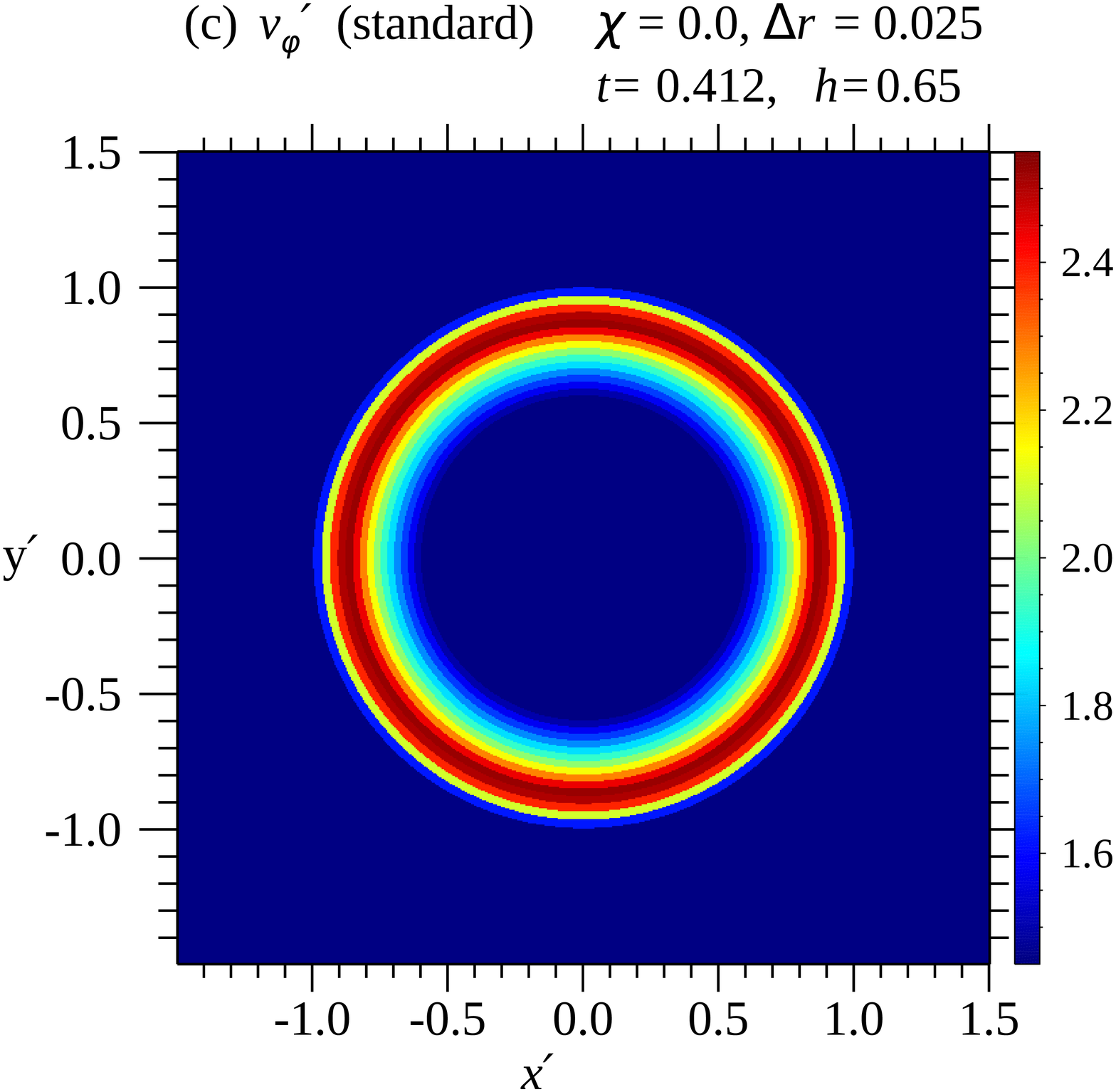}{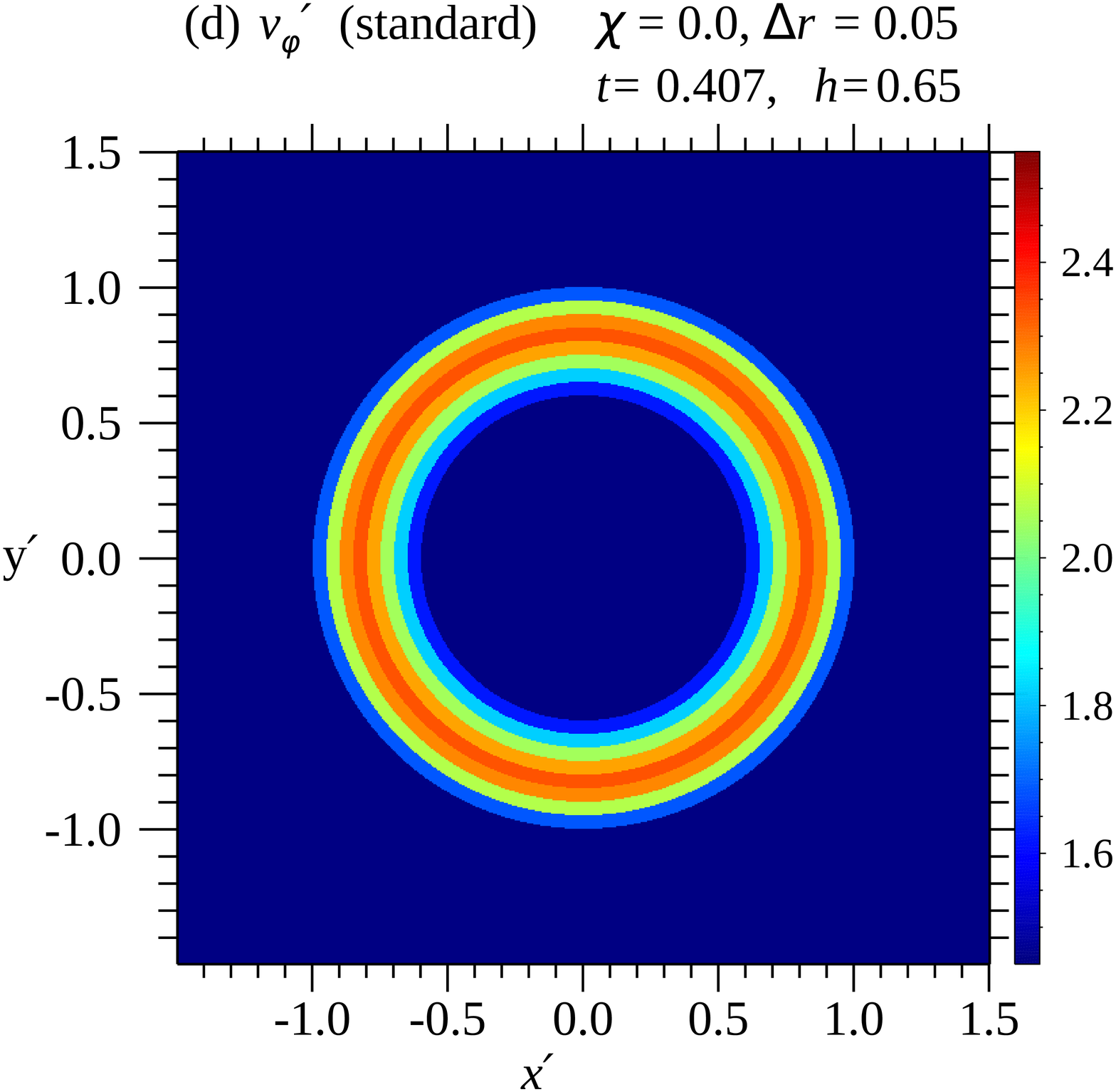}
\epsscale{1.0}
\caption{The color denotes the rotation velocity on a plane perpendicular to the rotation axis at $ t \simeq 0.41 $. The plane is 0.65 apart from the origin.  Panels (a) and (b) denote the
solutions of $ \chi = 0.9 $ obtained with the proposed and standard schemes, respectively.  
Panels (c) and (d) denote the solution of $ \chi= 0$ obtained with the current standard scheme in 2D with $ \Delta r = 0.025 $ and 0.05, respectively.  \label{outflow3}}
\end{figure}

Figure~\ref{outflow3} shows the rotation velocity in the plane perpendicular
to the rotation axis ($v_\varphi^\prime$) and above the origin by 0.65, i.e.,
near the bottom of the cavity. The $ x ^\prime $ axis runs in the plane of
$ \varphi = \lambda $, while the $ y ^\prime $ axis is perpendicular to both
the rotation axis and $ x ^\prime $-axis.
Panels (a) and (b) show the solution of $ \chi = 0.9 $ obtained with the proposed and standard schemes, respectively. Panel (c) shows that of $ \chi = 0 $ obtained with the standard scheme, while panel (d) is the same as panel (c) but with the lower resolution of $ \Delta r = 0.05$.  In these models, we performed 2D simulations in order to save computation time and 
to suppress numerical diffusion.
All the panels show the stage around $ t \simeq 0.41$.
When the model is symmetric around the $ z $-axis ($ \chi = 0 $), 
both the numerical schemes give almost the same solutions. The rotating speed $ v _\varphi ^\prime = 2.5 $ in the 2D model with the high-resolution run as shown in panel (c).  In the $\chi=0.9$ runs in panels (a) and (b), the overall profile becomes non-axisymmnetric because of the oblique expansion with respect to the coordinate system. Whereas a defect around $(x ^\prime,~y^\prime) = (0.8,~0.0) $ can be found in the standard scheme when the flow passed through the coordinate axis, $ r =0 $, we could obtain a smooth profile in the proposed scheme. Note that the peak rotation velocity is slightly lower in the models shown in panels (a), (b) and (d). This is due to the numerical diffusion and can be improved by improving the spatial resolution.

\subsection{Gas-disk Interaction around a Compact Object}

In this section, we show the simulation of an infalling cloudlet onto
a gas disk rotating around a compact object.  This test problem is 
motivated by  \citet{dullemond19,kuffmeier20}, who consider the capture 
of a fragment of molecular cloud by a young stellar object.
They propose a scenario in which the capture results in the FU Orionis burst.
A similar situation was also considered by \cite{kawashima17} for the delayed
brightening of Sgr A$^*$.  

The gravity is assumed to be as the following in this simulation,
\begin{eqnarray}
\bmv{g} & = & \left\{ 
\begin{array}{lcl}
\displaystyle - \frac{GM}{\left( r ^2 + z ^2 \right) ^{3/2}} \left( r \bmv{e} _r + z \bm{e} _z \right) & 
\mbox{for} & \sqrt{r ^2 + z ^2} \ge a \\
\displaystyle - \frac{GM}{a ^3} \left( r \bmv{e} _r + z \bm{e} _z \right) & \mbox{for} &
\sqrt{r ^2 + z ^2} < a \\
\end{array}
\right. .
\end{eqnarray}
This gravity is the same as that around a compact object except for
the small region around the origin, $ \sqrt{r ^2 + z ^2} < a $, where the gravity is
softened artificially.  We use the unit system, $ G M = 1 $ and $ a = 1 $.

We consider a cloudlet accreting onto a gas disk rotating around the origin.
The cloudlet and disk are surrounded by hot tenuous gas.  The density and pressure
distributions are described in Appendix A.  The specific heat ratio is set to be
$ \gamma = 1.05 $ to mimic an almost isothermal gas.

\begin{figure}
\plottwo{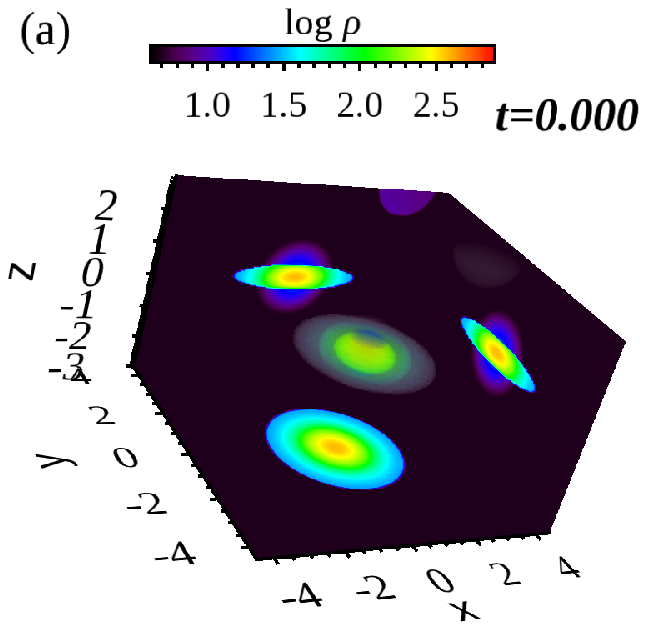}{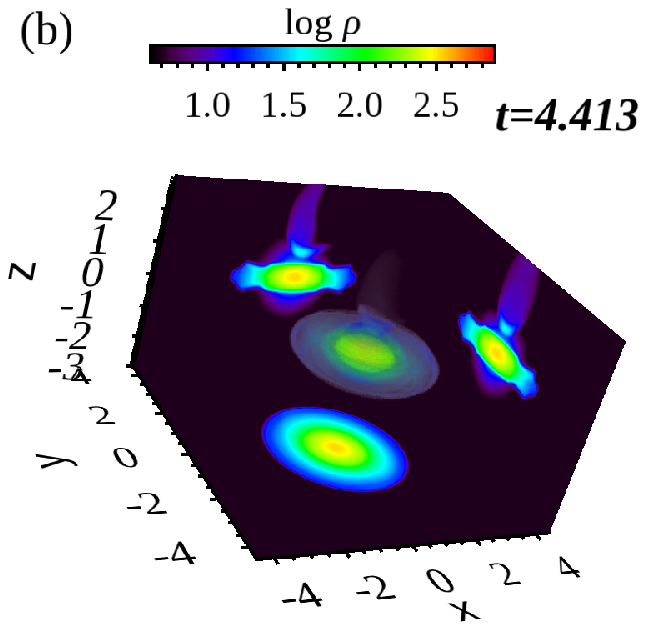}
\plottwo{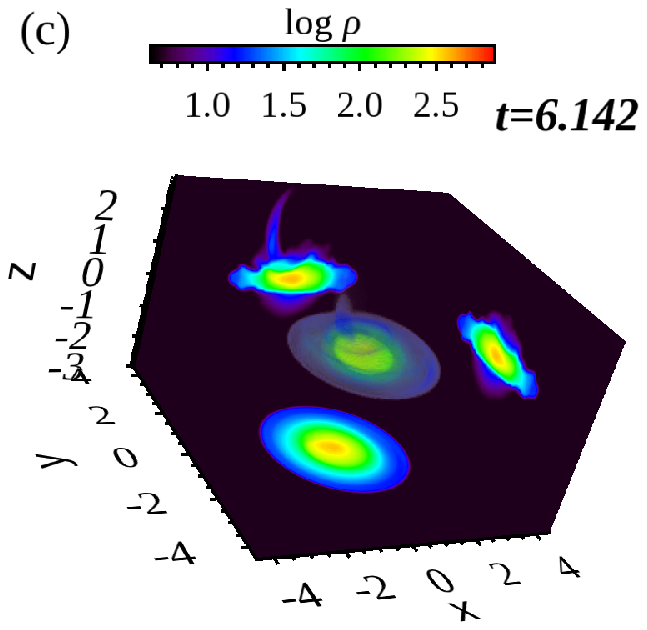}{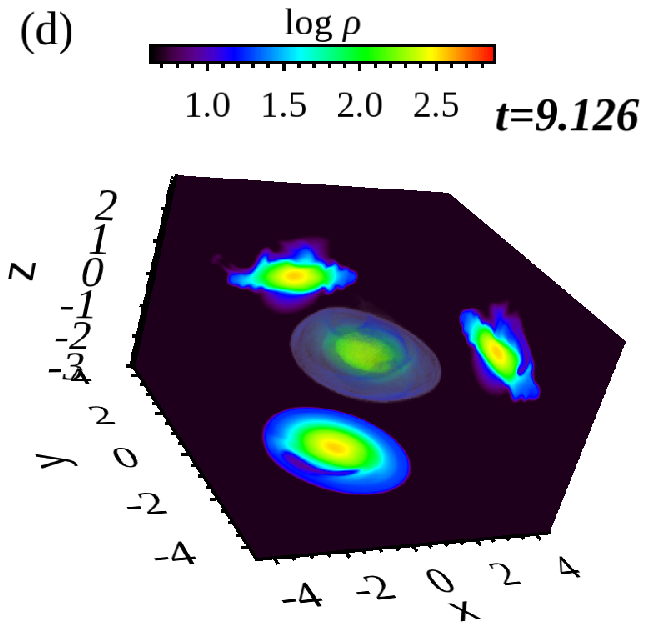}
\caption{A series of snapshots denote the infalling cloudlet onto 
a rotating gas disk surrounding a compact object.  Each panel denotes the
density by volume rendering and cross sections as in Figure \ref{outflow1}.
\label{accretion}}
\end{figure}

Figure \ref{accretion} shows the accretion of the cloudlet by a series of snapshots.  Each panel denotes the density distributions by volume rendering (center)
and three cross sections ($ x = 0 $, $ y = 0 $, and $ z = 0 $).  The upper left
panel shows the initial state.  The center of the cloudlet moves in the y = 0 plane.   
Panel (b) shows a stage in which the cloudlet passes the $ z $-axis.
Panel (c) shows a subsequent stage in which the cloudlet hits the
surface of the gas disk.  The cloudlet is absorbed into the rotating gas disk.  
We can find the impact of the infalling cloudlet in the panel (d).   
This example shows that our new scheme can handle this dynamic and 
nonaxisymmetric gas accretion onto a gas disk rotating around a compact object.

\section{Discussion}

As shown in the previous section, our new scheme is superior to the current standard scheme.
First, we can take a much longer time step since the angular resolution is only $ \Delta \varphi = \pi / 3 $.
Despite the low angular resolution, our new scheme achieves a higher accuracy for the free-stream problem, the Sod shock tube 
and rotating expanding gas sphere problems.  Our new scheme can suppress the disturbances
arising around the $ z $-axis, i.e., the apparent singularity, as shown in the expanding gas sphere
problem.

Among the tests shown in the previous section, the uniform flow test, i.e., the test for the free-stream preservation, is most fundamental and essential. A numerical solution of the shock tube problem does not converge to the exact one unless the scheme can solve a uniform flow without a numerical error. Remember that the density, velocity, and pressure are uniform in the exact solution in the period, $ t  > 0 $.  The situation is similar in the other test problems. It is important for a successful scheme to handle a uniform flow across the $ z $-axis.  Our scheme is designed to solve this problem perfectly.

{In the rest of this section, we discuss the validity of our approach by analyzing the new concepts adopted in our scheme.
They include the source term evaluation, the data reconstruction for numerical flux evaluation, considering the change 
in the unit vector, and correction factors.  We also discuss possible extension of our scheme to other system equations.} 

First, we review the hydrodynamic equations to justify our method of evaluating the apparent source term.
The inner product of Equation (\ref{hydroV2}) with $ \bmv{e} _r $, $ r \bmv{e} _\varphi $, and $ \bmv{e} _z $ 
provides Equation (\ref{hydro2}). 
Note that the third component of Equation (\ref{hydro2})
denotes not the conservation of the linear momentum but that of the angular momentum.
The latter is consistent with the pressure force derived from the numerical cell, and the azimuthal 
change in the pressure, $ \partial P / \partial \varphi $, denotes the torque.  Because 
$ ( \bmv{e} _r , r \bmv{e} _\varphi, \bmv{e} _z ) $ are the covariant unit vectors in
the cylindrical coordinates, we expect that we should discretize the conservation of generalized
momentum in curvilinear coordinates. The conservation law is expressed as follows,
\begin{eqnarray}
\frac{\partial}{\partial t} \left( \sqrt{g} \rho u _\mu \right) +
\frac{\partial}{\partial x ^\nu} \left( \sqrt{g} T _\mu {} ^\nu \right) & = & \frac{\partial g _{\alpha \beta}}{\partial x ^\mu}
T ^{\alpha \beta} ,
\end{eqnarray}
for the curvilinear coordinates defined by the following,
\begin{eqnarray}
ds ^2 & = & g _{\alpha \beta} dx ^\alpha dx ^\beta ,
\end{eqnarray}
where Einstein's notation is used.  The symbol, $ \sqrt{g} $, denotes the square root of the determinant
of the matrix, $ g _{\alpha\beta} $.

As mentioned in \S 2, numerical simulations are based on the conservation of linear momentum
rather than the conservation of angular momentum in numerical aerodynamics and other fields of computational
fluid dynamics, even when the numerical grid is based on the cylindrical coordinates.   
Such a formulation employs different vector bases for physical variables and geometry \citep{vinokur74}. 
The philosophy is clear, and the proposed scheme has been working well for simulations of the gas flow
around a body with complex boundaries.   However, this scheme has some drawbacks if applied to astrophysical
flows.   One drawback is the lack of the conservation of angular momentum as already mentioned.
Another drawback is the shape of the numerical cell: it should be a polygon so that each cell surface
is flat in the Cartesian coordinates \citep{vinokur89}.  Otherwise, the direction of the surface is
not well defined.  In short, this formulation is a special case of mesh-free decomposition of space.
Consequently, the radial cell surface does not coincide with the surface of constant radius, and it is not well matched with an adaptive angular resolution with the radius shown in Figure \ref{AMR}.  
The cell has an irregular shape at the coarse--fine grid interface.  Thus, we recommend our
formulation for simulations of astrophysical flows around a compact object.

We can apply the vector form described to the magnetohydrodynamic (MHD) equations, and the magnetic force is expressed as follows,
\begin{eqnarray}
\frac{\bmv{j} \times \bmv{B}}{c} & = & \frac{1}{4 \pi} \bmv{\nabla} \cdot 
\left( \bmv{BB} - \frac{\bmv{B}\cdot\bmv{B}}{2}   \right) , \\
\left( \frac{\bmv{j} \times \bmv{B}}{c} \right) _r & = & \frac{1}{8 \pi} \left\{
\frac{\partial}{\partial r} \left( B _r ^2 - B _\varphi ^2 - B _z ^2 \right) + \frac{1}{r} 
\frac{\partial}{\partial \varphi}  \left( B _r B _\varphi \right) 
+ \frac{\partial}{\partial z} \left( B _r B _z \right) - \frac{B _\varphi ^2}{r}  \label{jB2}
\right\},
\end{eqnarray}
where $ c $ denotes the light speed.  The hoop stress, $ B _\varphi ^2 / r $, in
Equation (\ref{jB2}) is similar to the centrifugal force \cite[see, e.g.,][for the source term in the MHD equations in the cylindrical coordinates]{skinner10}.  We think that this term
should be evaluated on the cell surfaces in the azimuthal direction, although
it is evaluated at the cell center in the current standard scheme.

The presented arguments are based on formality and aesthetics.  However, we think our
treatment of the centrifugal force has practical merit to improve the robustness.
Because we use the average of the cell surface values, the difference from the cell center 
value is proportional to the second derivative.  In other words, the additional numerical diffusion
suppresses disturbances appearing around the $ z $-axis in our scheme.
We think that the current standard discretization equations contain
negative diffusion as a truncation error in the centrifugal force.  
Negative diffusion is known to amplify small-scale disturbances. This interpretation is consistent with
our experiences that we often find small disturbances around the $ z $-axis in the numerical simulations
performed on the cylindrical coordinates. The disturbances around the pole ($ \theta = 0 $) in the numerical
solutions performed in spherical coordinates are likely to be of the same origin.

Another new feature of our scheme is the data reconstruction stage in MUSCL, and we have used the
local Cartesian coordinates to evaluate the velocity.   We have constrained not the radial and
azimuthal components, $ v _r $ and $ v _\varphi $, but the normal
and tangential components of the velocity, $ v _{\rm n} $ and $ v _{\rm t} $, to be monotonic.  
The principle of monotonicity preserving is ascribed to the fact that the wave amplitude is
a Riemann invariant and does not change during the propagation.  Thus, we should constrain
the wave amplitude to be monotonic in principle.  We often constrain the primitive
variables such as $ \rho $, $ P $, and $\bmv{v} $ in order
to save extra computational cost for computing wave amplitudes.  This simple and easy
method works in the Cartesian coordinates but may not in the cylindrical coordinates.
In the latter, a part of the changes in the radial and azimuthal components of the velocity may be
ascribed to those in the unit vector. Such changes are eliminated in the covariant derivative
\citep[see, e.g.][]{mitra09}.
We should use the covariant derivative to check the monotonicity of the velocity.
This procedure is implemented in the Pencil Code \citep{pencil18}.
As noted in \cite{mitra09} and the manual of the Pencil Code, we should also use the covariant 
derivative of the magnetic field in the MHD to exclude the change because of the curvature of the 
coordinates.  The velocity and magnetic field in the local Cartesian grid are easier to compute than
the covariant derivative; yet, the accuracy is of a higher order.

The reduction of the angular resolution around the $ z $-axis is not quite new.  It has already been
adopted in the literature \citep[see, e.g.,][]{liska18}.   However, our scheme is quite different 
because we can arrange the angular resolution almost seamlessly.   The aspect ratio is set to be nearly uniform
in the test examples shown in this paper. However, we can take a higher angular resolution
in the middle if needed, and the angular resolution can be adjusted systematically, as in AMR.

Although our mesh refinement is similar to AMR, we have modified the data reconstruction
across the coarse--fine grid boundary.  As mentioned in \S 3, the azimuthal width of the radial
cell surface is set equal to the minimum azimuthal widths among the numerical cells used
in the data construction.   As a result, the azimuthal width of the radial cell surface can be
smaller than those of the adjacent cells.  In contrast, we have not considered
the azimuthal gradient in the data reconstruction even when the cell considered is wider
than that of the cell surface.   The computation of the radial flux across the coarse--fine
boundary is simplified in our scheme.   

Among our improvements, the correction factor to the azimuthal cell surface may be
most arguable, and it is introduced to eliminate the truncation error for uniform flow.
However, the truncation error comes from the surface integral in the $ r $-direction.  
The unit vector, $ \bmv{e} _r $, is not uniform on the radial cell surface; hence, the following inequality is obtained,
\begin{eqnarray}
\int _{\varphi _{j-1/2}} ^{\varphi _{j+1/2}} \int _{z _{k-1/2}} ^{z _{k+1/2}} \bmv{e} _r r_{i+1/2} d\varphi d z
& \ne & r _{i+1/2} \Delta \varphi _j \Delta z _k \bmv{e} _r (\varphi _j) .
\end{eqnarray}
This truncation error is compensated by the correction factor applied to the azimuthal
direction. One might question why we do not apply the correction factor to the radial direction, but it was difficult for us to find such a correction factor.   As seen in Table 
\ref{Scorrection}, while it should depend on the angular resolution, it should also have the
same value on the coarse--fine grid boundary to guarantee the conservation.
We are afraid that such a correction factor may not exist.

At present, we do not know whether we can eliminate truncation error for a uniform
flow for given curved linear coordinates.  The truncation error is of the second-order
small quantities proportional to the curvature of the coordinates.   If we can reduce
it to the third or higher small quantities, it will be beneficial, and it is a challenging problem
to find such a discretization form for curved linear coordinates in general.

\acknowledgments

This work is supported by JSPS KAKENHI Grant Number JP19K03906.  We thank an anonymous
reviewer for their suggestions and comments.
We also thank Dinshaw Balsara, Akira Mizuta, Ryoji Matsumoto, and Tomoya Takiwaki for
providing valuable information in the literature.

%

\appendix

\section{Cloudlet Accretion onto a Rotating Gas Disk}

In the model of gas-disk interaction around a compact object, we consider a tenuous surrounding gas.  The tenuous gas is assumed to be isothermal and in hydrostatic equilibrium.
The density and pressure distributions are expressed as follows,
\begin{eqnarray}
\rho _{\rm h} (r, z) & = & \rho _0 
\exp \left[ - \frac{\Phi (r, z)}{T _{\rm h}} \right] , \\
P _{\rm h} (r, z) & = & \rho _{\rm h} (r, z) T _{\rm h} . 
\end{eqnarray}
The gas temperature is set to be $ T _{\rm h} = GM / (2a) $.   

The gas in the disk is assumed to be supported mainly by the centrifugal force against the gravity.  
The rotation velocity is expressed as follows,
\begin{eqnarray}
v _\varphi (r, z) & = & \Omega _{\rm K} r
\sqrt{\left( 1 - \frac{GM}{a T _{\ell}} \right) \left( 1 + \beta ^2 \right)}  , \\ 
\Omega _{\rm K} & = & \left\{ 
\begin{array}{lll}
\left( \displaystyle \frac{GM}{a ^3} \right) ^{1/2} & \mbox{for} \; & r ^2 + z ^2 \le a ^2 \\
\left[ \displaystyle \frac{GM}{(r ^2 + z ^2) ^{3/2}} \right] ^{1/2} & \mbox{for} \; & r ^2 + z ^2 > a ^2 \\
\end{array}
\right. ,
\end{eqnarray}
where the gas temperature and parameter are set to be $ T _{\ell} = GM/(10a) $ and $ \beta = 1/5$, 
respectively.   
The disk has the radius, $ r _{\rm disk} = 2 a $, and the half thickness,
\begin{eqnarray}
H & = & \beta \sqrt{ a ^2 + r ^2} ,
\end{eqnarray}
as a function of $ r $.  The gas disk is assumed to be isothermal at the temperature, $ T _{\rm \ell} $,
and in the hydrostatic equilibrium in the vertical direction.   The radial and vertical velocities are set
to vanish, $  v _r = v _z = 0 $.  
The gas pressure on the disk surface is
given by $ P _{\rm h} [r, H] $.

The infalling cloudlet is set in the following region,
\begin{eqnarray}
\left( r \cos \varphi - 3 a \right) ^2 + r ^2 \sin ^2 \varphi + \left( z - \sqrt{10} a \right) ^2 \le a ^2 .
\end{eqnarray}
It is assumed to have the same pressure as that of the surrounding ambient gas but to
have by a factor 5 higher density and hence by a factor 5 lower temperature.
The velocity is set to be as follows,
\begin{eqnarray}
v _r (r, z) & = & - \sqrt{\displaystyle \frac{20 G M}{33a}} , \\
v _\varphi (r, z) & = & \sqrt{\displaystyle \frac{2GM}{33a}} , \\
v _z (r, \varphi) & = & 0.
\end{eqnarray}
This infalling cloudlet is designed to flow across the $ z $-axis so we can assess the validity of our
scheme.   This model is not intended to simulate a more realistic situation.




\end{document}